\documentclass[11pt]{article}
\usepackage{graphicx,psfrag,latexsym,amsmath,amssymb,amsfonts,
	epsf,epsfig,cite,mathrsfs,subfigure,url,array,dsfont}

\newcommand{\dotle} {\mbox{$\:\stackrel{\centerdot}{\le}\:$}}

\newcommand{\calC}{{\cal C}}
\newcommand{\calD}{{\cal D}}
\newcommand{\calE}{{\cal E}}
\newcommand{\calF}{{\cal F}}

\newcommand{\calM}{{\cal M}}

\newcommand{\calX}{{\cal X}}
\newcommand{\calY}{{\cal Y}}

\newcommand{\bx}{{\mathbf x}}

\newcommand{\by}{{\mathbf y}}

\newcommand{\tp}{\tilde{p}}

\newcommand{\tby}{\tilde{\mathbf y}}

\newcommand{\scrP}{\mathscr{P}}
\newcommand{\scrW}{\mathscr{W}}

\def\eE{{\Bbb E}}

\def\rR{{\Bbb R}}

\newcommand{\Ni}{N_{\mathrm{i}}}
\newcommand{\Ei}{E_{\mathrm{i}}}

\newtheorem{theorem}{Theorem}[section]
\newtheorem{lemma}[theorem]{Lemma}
\newtheorem{proposition}[theorem]{Proposition}
\newtheorem{corollary}[theorem]{Corollary}
\renewcommand{\theequation}{\arabic{section}.\arabic{equation}}
\newcommand{\Section}[1]{\section{#1}
\setcounter{equation}{0}}

\begin{document}

\author{Pierre Moulin \\
	ECE Department, Beckman Inst., and Coordinated Science Lab. \\
	University of Illinois at Urbana-Champaign \\
	Urbana, IL, 61801, USA \\
	Email: {\tt moulin@ifp.uiuc.edu} 
	\thanks{This work was supported by NSF under grant CCF 06-35137
	and by DARPA under the ITMANET program.
	A 5-page version of this paper was submitted to ISIT'08.}
	}

\title{A Neyman-Pearson Approach \\ to Universal Erasure and List Decoding}

\maketitle

\begin{abstract}
When information is to be transmitted over an unknown, possibly unreliable channel,
an erasure option at the decoder is desirable.
Using constant-composition random codes, we propose a generalization of
Csisz\'{a}r and K\"{o}rner's Maximum Mutual Information decoder with erasure option
for discrete memoryless channels.
The new decoder is parameterized by a {\em weighting function} that is
designed to optimize the fundamental tradeoff between undetected-error and
erasure exponents for a compound class of channels.
The class of weighting functions may be further enlarged to optimize a similar
tradeoff for list decoders --- in that case, undetected-error probability
is replaced with average number of incorrect messages in the list.
Explicit solutions are identified.

The optimal exponents admit simple expressions
in terms of the sphere-packing exponent, at all rates below capacity.
For small erasure exponents, these expressions coincide with those derived by Forney (1968) 
for symmetric channels, using Maximum a Posteriori decoding. Thus for those channels
at least, ignorance of the channel law is inconsequential.
Conditions for optimality of the Csisz\'{a}r-K\"{o}rner rule and of the simpler
empirical-mutual-information thresholding rule are identified.
The error exponents are evaluated numerically for the binary symmetric channel.
\\[12pt]

\end{abstract}

\newpage

\Section{Introduction}

Universal decoders have been studied extensively in the information theory literature
as they are applicable to a variety of communication problems where the channel
is partly or even completely unknown \cite{Csiszar81,Lapidoth98}.
In particular, the Maximum Mutual Information (MMI) decoder provides
universally attainable error exponents for random constant-composition codes
over discrete memoryless channels (DMCs). In some cases,
incomplete knowledge of the channel law is inconsequential
as the resulting error exponents are the same as those for 
maximum-likelihood decoders which know the channel law in effect.

It is often desirable to provided the receiver with
an erasure option that can be exercised when the received data are deemed 
unreliable. For fixed channels, Forney \cite{Forney68} derived the decision rule
that provides the optimal tradeoff between the erasure and undetected-error probabilities,
analogously to the Neyman-Pearson problem for binary hypothesis testing.
Forney used the same framework to optimize the performance of list decoders;
the probability of undetected errors is then replaced by the expected number of
incorrect messages on the list. The size of the list is a random variable which
equals 1 with high probability when communication is reliable.

For unknown channels, the problem of decoding with erasures was
considered by Csisz\'{a}r and K\"{o}rner \cite{Csiszar81}. They derived attainable
pairs of undetected-error and erasure exponents for any DMC.
Their work was later extended by Telatar and Gallager \cite{Telatar94}.
However neither \cite{Csiszar81} nor \cite{Telatar94} indicated whether true
universality is achievable, i.e., whether the exponents match Forney's exponents.
Also they did not indicate whether their error exponents might be optimal
in some weaker sense. The problem was recently revisited by Merhav and Feder \cite{Merhav07},
using a competitive minimax approach. The analysis of \cite{Merhav07} yields lower
bounds on a certain fraction of the optimal exponents.
It is suggested in \cite{Merhav07} that true universality might generally not be
attainable, which would represent a fundamental difference with ordinary decoding.

This paper considers decoding with erasures for the compound DMC, with two goals in mind.
The first is to construct a broad class of decision rules that can be
optimized in an asymptotic Neyman-Pearson sense, analogously to universal hypothesis
testing \cite{Hoeffding65,Tusnady77,Zeitouni91}.
The second is to investigate the universality properties of the receiver,
in particular conditions under which the exponents coincide with Forney's exponents.
We first solve the problem of variable-size list decoders because it is simpler,
and the solution to the ordinary problem of size-1 lists follows directly.
We establish conditions under which our error exponents match Forney's exponents.

Following background material in Sec.~\ref{sec:decoding},
the main results are given in Secs.~\ref{sec:F-MMI}---\ref{sec:optimal-F}.
We also observe that in some problems the compound DMC approach is overly rigid and
pessimistic. For such problems we present in Sec.~\ref{sec:relative} a simple and
flexible extension of our method based on the relative minimax principle.
In Sec.~\ref{sec:BSC} we apply our results to a class of Binary Symmetric
Channels (BSC), which yields easily computable and insightful formulas.
The paper concludes with a brief discussion in Sec.~\ref{sec:discussion}.
The proofs of the main results are given in the appendices.

\subsection{Notation}

We use uppercase letters for random variables,
lowercase letters for individual values, and boldface fonts for sequences.
The probability mass function (p.m.f.) of a random variable $X \in \calX$
is denoted by $p_X = \{ p_X(x), \, x \in \calX \}$, the probability 
of a set $\Omega$ under $p_X$ by $P_X(\Omega)$,
and the expectation operator by $\eE$.
Entropy of a random variable $X$ is denoted by $H(X)$,
and mutual information between two random variables $X$ and $Y$
is denoted by $I(X;Y) = H(X) - H(X|Y)$, or by $I(p_{XY})$
when the dependency on $p_{XY}$ should be explicit.
The Kullback-Leibler divergence between two p.m.f.'s $p$ and $q$ is denoted by $D(p||q)$. 
All logarithms are in base 2.
We denote by $f'$ the derivative of a function $f$.

Denote by $p_\bx$ the type of a sequence $\bx \in \calX^N$
($p_\bx$ is an empirical p.m.f. over $\calX$) and by $T_\bx$ the type class 
associated with $p_\bx$, i.e., the set of all sequences of type $p_\bx$.
Likewise, denote by $p_{\bx\by}$ the joint type of a pair of sequences
$(\bx, \by) \in \calX^N \times \calY^N$
(a p.m.f. over $\calX \times \calY$) and by $T_{\bx\by}$ the type class
associated with $p_{\bx\by}$, i.e., the set of all sequences
of type $p_{\bx\by}$.
The conditional type $p_{\by|\bx}$ of a pair of sequences ($\bx, \by$)
is defined as $p_{\bx\by}(x,y)/p_{\bx}(x)$ for all $x \in \calX$
such that $p_{\bx}(x) > 0$. 
The conditional type class $T_{\by|\bx}$ is the set of
all sequences $\tby$ such that $(\bx, \tby) \in T_{\bx\by}$.
We denote by $H(\bx)$ the entropy of the p.m.f. $p_{\bx}$ and by $H(\by|\bx)$
and $I(\bx;\by)$ the conditional entropy and the mutual information
for the joint p.m.f. $p_{\bx\by}$, respectively.
Recall that \cite{Csiszar81}
\begin{eqnarray}
   (N+1)^{- |\calX|} \,2^{NH(\bx)} \le & |T_{\bx}| & \le 2^{NH(\bx)} ,	\label{eq:type-size1} \\
   (N+1)^{- |\calX| \,|\calY|} \,2^{NH(\by|\bx)} 
	\le & |T_{\by|\bx}| & \le 2^{NH(\by|\bx)} . 						\label{eq:type-size2}
\end{eqnarray}

We let $\scrP_X$ and $\scrP_X^{[N]}$ represent the set of all 
p.m.f.'s and empirical p.m.f.'s, respectively, for a random variable $X$.
Likewise, $\scrP_{Y|X}$ and $\scrP_{Y|X}^{[N]}$ denote
the set of all conditional p.m.f.'s and all empirical conditional p.m.f.'s, 
respectively, for a random variable $Y$ given $X$.
The notation $f(N) \sim g(N)$ denotes asymptotic equality:
$\lim_{N \rightarrow \infty} \frac{f(N)}{g(N)} = 1$.
The shorthands $f(N) \doteq g(N)$ and $f(N) \dotle g(N)$ 
denote equality and inequality on the exponential scale:
$\lim_{N \rightarrow \infty} \frac{1}{N} \ln \frac{f(N)}{g(N)} = 0$ and
$\lim_{N \rightarrow \infty} \frac{1}{N} \ln \frac{f(N)}{g(N)} \le 0$,
respectively.
We denote by $\mathds1_{\{ x \in \Omega \}}$ the indicator function of a set $\Omega$
and define $|t|^+ \triangleq \max(0,t)$ and  $\exp_2(t) \triangleq 2^t$.
We adopt the notational convention that the minimum of a function over an empty set
is $+ \infty$.

The function-ordering notation $F \preceq G$ indicates that $F(t) \le G(t)$ for all $t$.
Similarly, $F \succeq G$ indicates that $F(t) \ge G(t)$ for all $t$.

\Section{Decoding with Erasure and List Options}
\label{sec:decoding}
\setcounter{equation}{0}

\subsection{Maximum-Likelihood Decoding}
\label{sec:ML}

In his 1968 paper\cite{Forney68}, Forney studied the following erasure/list decoding problem.
A length-$N$, rate-$R$ code $\calC = \{ \bx(m), \,m \in \calM \}$ is selected, where
$\calM = \{ 1, 2, \cdots, 2^{NR} \}$ is the message set and each codeword  $\bx(m) \in \calX^N$.
Upon selection of a message $m$, the corresponding $\bx(m)$
is transmitted over a DMC $p_{Y|X}~:~\calX \to \calY$.
A set of decoding regions $\calD_m \subseteq \calY^N, \, m \in \calM$,
is defined, and the decoder returns $\hat{m} = g(\by)$ if and only if $\by \in \calD_m$.
For ordinary decoding, $\{ \calD_m,\, m \in \calM \}$ form a partition of $\calY^N$.
When an erasure option is introduced, the decision space
is extended to $\calM \cup \emptyset$, where $\emptyset$ denotes the erasure symbol.
The erasure region $\calD_\emptyset$ is the complement of
$\cup_{m \in \calM} \calD_m$ in $\calY^N$.
An undetected error arises if $m$ was transmitted but $\by$ lies in the
decoding region of some other message $i \ne m$. This event is given by
\begin{equation}
   \calE_{\mathrm i} = \left\{ (m,\by) ~:~ \by \in \bigcup_{i \in \calM \setminus \{m\}} \calD_i \right\} ,
\label{eq:u}
\end{equation}
where the subscript i stands for ``incorrect message''. Hence
\begin{eqnarray}
   Pr[\calE_{\mathrm i}] & = & \frac{1}{|\calM|} \sum_{m \in \calM}
	\,\sum_{\by \in \bigcup_{i \in \calM \setminus \{m\}} \calD_i} p_{Y|X}^N(\by|\bx(m)) \nonumber \\
	& = & \frac{1}{|\calM|} \sum_{m \in \calM} \,\sum_{i \in \calM \setminus \{m\}}
	\,\sum_{\by \in \calD_i} p_{Y|X}^N(\by|\bx(m)) ,
\label{eq:Pr-u}
\end{eqnarray}
where the second equality holds because the decoding regions are disjoint.

The erasure event is given by
\[ \calE_\emptyset = \{ (m,\by) ~:~ \by \in \calD_\emptyset \} \]
and has probability
\begin{equation}
   Pr[\calE_\emptyset] = \frac{1}{|\calM|} \sum_{m \in \calM} \sum_{\by \in \calD_\emptyset}
	p_{Y|X}^N(\by|\bx(m)) .
\label{eq:Pr-e}
\end{equation}
The total error event is given by $\calE_{err} = \calE_{\mathrm i} \cup \calE_\emptyset$.
The decoder is generally designed so that $Pr[\calE_{\mathrm i}] \ll Pr[\calE_\emptyset]$,
so $Pr[\calE_{err}] \approx Pr[\calE_\emptyset]$. 

Analogously to the Neyman-Pearson problem, one wishes to design the decoding
regions to obtain an optimal tradeoff between $Pr[\calE_{\mathrm i}]$ and $Pr[\calE_\emptyset]$.
Forney proved the following class of decision rules is optimal:
\begin{equation}
   g_{ML}(\by) = \left\{ \begin{array}{ll}
				\hat{m} & :~ \mathrm{if~} p_{Y|X}^N(\by|\bx(\hat{m}))
					> e^{NT} \,\underset{i \ne \hat{m}}{\sum} p_{Y|X}^N(\by|\bx(i))  \\
				\emptyset & :~\mathrm{else}
				\end{array} \right.
\label{eq:decision-forney}
\end{equation}
where $T \ge 0$ is a free parameter trading off $Pr[\calE_{\mathrm i}]$ against
$Pr[\calE_\emptyset]$. The nonnegativity constraint on $T$ ensures that
$\hat{m}$ is uniquely defined for any given $\by$.
There is no other decision rule that yields simultaneously a lower
value for $Pr[\calE_{\mathrm i}]$ and for $Pr[\calE_\emptyset]$.

A conceptually simple (but suboptimal) alternative to (\ref{eq:decision-forney}) is
\begin{equation}
   g_{ML,2}(\by) = \left\{ \begin{array}{ll}
				\hat{m} & :~ \mathrm{if~} p_{Y|X}^N(\by|\bx(\hat{m}))
					> e^{NT} \,\underset{i \ne \hat{m}}{\max} \,p_{Y|X}^N(\by|\bx(i))  \\
				\emptyset & :~\mathrm{else}
				\end{array} \right.
\label{eq:decision-forney-2}
\end{equation}
where the decision is made based on the two highest likelihood scores.

If one chooses $T < 0$, there is generally more than one value of $\hat{m}$ that satisfies
(\ref{eq:decision-forney}), and $g_{ML}$ may be viewed as a list decoder
that returns the list of all such $\hat{m}$. Denote by $\Ni$ the number of incorrect
messages on the list. Since the decoding regions $\{\calD_m, \,m \in \calM\}$
overlap, the average number of incorrect messages in the list,
\begin{equation}
   \eE[\Ni] = \frac{1}{|\calM|} \sum_{m \in \calM} \,\sum_{i \in \calM \setminus \{m\}}
	\,\sum_{\by \in \calD_i} p_{Y|X}^N(\by|\bx(m)) ,
\label{eq:ENi}
\end{equation}
no longer coincides with $Pr[\calE_{\mathrm i}]$ in (\ref{eq:Pr-u}).

For the rule (\ref{eq:decision-forney}) applied to symmetric channels, Forney showed
that the following error exponents are achievable for all $\Delta$ such that
$R^{conj} \le R+\Delta \le C$:
\begin{eqnarray}
   \Ei(R,\Delta) & = & E_{sp}(R) + \Delta \nonumber \\
	E_\emptyset(R,\Delta) & = & E_{sp}(R+\Delta)
\label{eq:E-forney}
\end{eqnarray}
where $E_{sp}(R)$ is the sphere-packing exponent, and
$R^{conj}$ is the conjugate rate, defined as the rate for which the slope
of $E_{sp}(\cdot)$ is the inverse of the slope at rate $R$:
\[ E_{sp}'(R^{conj}) = \frac{1}{E_{sp}'(R)} . \]
The exponents of (\ref{eq:E-forney}) are achieved using independent and identically
distributed (i.i.d.) codes.

\subsection{Universal Decoding}
\label{sec:erasure}

When the channel law $p_{Y|X}$ is unknown, maximum-likelihood decoding cannot be used.
For constant-composition codes with type $p_X$, the MMI decoder takes the form
\begin{equation}
   g_{MMI}(\by) = \arg\max_{i \in \calM} \,I(\bx(i);\by) .
\label{eq:MMI}
\end{equation}
Csisz\'{a}r and K\"{o}rner \cite[p.~174---178]{Csiszar81} extended
the MMI decoder to include an erasure option, using the following decision rule:
\begin{equation}
   g_{\lambda,\Delta}(\by) = \left\{ \begin{array}{ll} \hat{m}
				& :~ \mathrm{if~} I(\bx(\hat{m});\by) > R + \Delta 
					+ \lambda \;\underset{i \ne \hat{m}}{\max} |I(\bx(i);\by) - R|^+  \\
				\emptyset & :~\mathrm{else}
				\end{array} \right.
\label{eq:decision-CK}
\end{equation}
where $\Delta \ge 0$ and $\lambda > 1$.
They derived the following error exponents for the resulting undetected-error
and erasure events:
\[ \{ E_{r,\lambda}(R,p_X,p_{Y|X}) + \Delta ,\; E_{r,1/\lambda}(R+\Delta,p_X,p_{Y|X}) \} ,
		\quad \forall p_{Y|X} \]
where
\[ E_{r,\lambda}(R,p_X,p_{Y|X}) = \min_{\tp_{Y|X}} \{ D(\tp_{Y|X}\|p_{Y|X} |p_X) 
	+ \lambda |I(p_X, \tp_{Y|X}) - R|^+ \} . \]
While $\Delta$ and $\lambda$ are tradeoff parameters,
they did not mention whether the decision rule (\ref{eq:decision-CK})
satisfies any Neyman-Pearson type optimality criterion. 

A different approach was recently proposed by Merhav and Feder \cite{Merhav07}.
They raised the possibility that the achievable pairs of undetected-error and erasure
exponents might be smaller than in the known-channel case and proposed
a decision rule based on the competitive minimax principle.
This rule is parameterized by a scalar parameter $0 \le \xi \le 1$ which represents
a fraction of the optimal exponents (for the known-channel case) that their
decoding procedure is guaranteed to achieve. Decoding
involves explicit maximization of a cost function over the compound DMC
family, analogously to a Generalized Likelihood Ratio Test (GLRT).
The rule coincides which the GLRT when $\xi=0$, but the choice
of $\xi$ can be optimized. They conjectured that the highest achievable $\xi$
is lower than 1 in general, and derived a computable lower bound on that value.

\Section{$\calF$--MMI Class of Decoders}
\label{sec:F-MMI}
\setcounter{equation}{0}

\subsection{Decoding Rule}

Assume that random constant-composition codes with type $p_X$ are used, and that
the DMC $p_{Y|X}$ belongs to a {\em connected subset} $\scrW$ of $\scrP_{Y|X}$.
The decoder knows $\scrW$ but not which $p_{Y|X}$ is in effect.

Analogously to (\ref{eq:decision-CK}), our proposed decoding rule is a test based on
the empirical mutual informations for the two highest-scoring messages.
Let $\calF$ be the class of continuous, nondecreasing functions
$F~:~[-R, \,H(p_X)-R] \to \rR$. 
The decision rule indexed by $F \in \calF$ takes the form 
\begin{equation}
  g_F(\by) = \left\{ \begin{array}{ll} \hat{m}
					& :~ \mathrm{if~} I(\bx(\hat{m});\by) > R  
						+ \underset{i \ne \hat{m}}{\max} \,F(I(\bx(i);\by) - R)  \\
					\emptyset & :~\mathrm{else.}
					\end{array} \right.
\label{eq:decision}
\end{equation}
Given a candidate message $\hat{m}$, the function $F$ weighs the score of the best
competing codeword.
Since $0 \le I(\bx(i);\by) \le H(p_X)$, all values of $F(t)$ outside the range
$[-R, H(p_X)-R]$ are equivalent in terms of the decision rule (\ref{eq:decision}).

%

The choice $F(t)=t$ results in the MMI decoding rule (\ref{eq:MMI}), and
\begin{equation}
   F(t) = \Delta + \lambda\,|t|^+
\label{eq:F-CK}
\end{equation}
(two-parameter family of functions)
results in the Csisz\'{a}r-K\"{o}rner rule (\ref{eq:decision-CK}).

One may further require that $F(t) \ge t$ to guarantee that
$\hat{m} = \arg\max_i \,I(\bx(i);\by)$, as can be verified by direct substitution
into (\ref{eq:decision}). In this case, the decision is whether the decoder should
output the highest-scoring message or output an erasure decision.

When the restriction $F(t) \ge t$ is not imposed,
the decision rule (\ref{eq:decision}) is ambiguous because more
than one $\hat{m}$ could satisfy the inequality in (\ref{eq:decision}).
Then (\ref{eq:decision}) may be viewed as a list decoder that returns the list
of all such $\hat{m}$, similarly to (\ref{eq:decision-forney}).



The Csisz\'{a}r-K\"{o}rner decision rule parameterized by $F$ in (\ref{eq:F-CK})
is nonambiguous for $\lambda \ge 1$. Note there is an error in
Theorem~5.11 and Corollary~5.11A of \cite[p.~175]{Csiszar81},
where the condition $\lambda > 0$ should be replaced with $\lambda \ge 1$
\cite{Csiszar07}.

In the limit as $\lambda \downarrow 0$, (\ref{eq:F-CK}) leads to the simple decoder
that lists all messages whose empirical mutual information score exceeds $R+\Delta$.
If a list decoder is not desired, a simple variation on (\ref{eq:F-CK}) 
when $0 < \lambda < 1$ is
\[ F(t) = \left\{ \begin{array}{ll}
	\Delta + \lambda\,|t|^+ & :~t \le \frac{\Delta}{1-\lambda} \\
	t & :~\mathrm{else.}
	\end{array} \right.
\]

It is also worth noting that the function $F(t) = \Delta + t$ may be thought
of as an empirical version of Forney's suboptimal decoding rule
(\ref{eq:decision-forney-2}), with $T=\Delta$. Indeed, using the identity
$I(\bx;\by) = H(\by) - H(\by|\bx)$ and viewing the negative empirical equivocation
\[  - H(\by|\bx) = \sum_{x,y} p_{\bx\by}(x,y) \ln p_{\by|\bx}(y|x) \]
as an empirical version of the normalized loglikelihood
\[ \frac{1}{N} \ln p_{Y|X}^N(\by|\bx) = \sum_{x,y} p_{\bx\by}(x,y) \ln p_{Y|X}(y|x) , \]
we may rewrite (\ref{eq:decision-forney-2}) and (\ref{eq:decision}) respectively as
\begin{equation}
   g_{ML,2}(\by) = \left\{ \begin{array}{ll}
				\hat{m} & :~ \mathrm{if~} \frac{1}{N} \ln p_{Y|X}^N(\by|\bx(\hat{m}))
					> T + \,\underset{i \ne \hat{m}}{\max} \,\frac{1}{N} \ln p_{Y|X}^N(\by|\bx(i))  \\
				\emptyset & :~\mathrm{else}
				\end{array} \right.
\label{eq:decision-eqv1}
\end{equation}
and
\begin{equation}
  g_F(\by) = \left\{ \begin{array}{ll} \hat{m}
					& :~ \mathrm{if~} -H(\by|\bx(\hat{m})) > \Delta  
						+ \underset{i \ne \hat{m}}{\max} \,[-H(\by|\bx(i))]  \\
					\emptyset & :~\mathrm{else.}
					\end{array} \right.
\label{eq:decision-eqv2}
\end{equation}
While this observation does not imply $F(t) = \Delta + t$
is an optimal choice for $F$, one might intuitively expect optimality in some regime.

\subsection{Error Exponents}

For a random-coding strategy using constant-composition codes with
type $p_X$, the expected number of incorrect messages on the list, $\eE[\Ni]$,
and the erasure probability, $Pr[\calE_\emptyset]$, may be viewed
as functions of $R$, $p_X$, $p_{Y|X}$, and $F$. A pair
$\{\Ei(R,p_X,p_{Y|X},F), E_\emptyset (R,p_X,p_{Y|X},F)\}$ of
incorrect-message and erasure exponents is said to be universally attainable
for such codes over $\scrW$ if the expected number of incorrect messages on the list
and the erasure probability satisfy
\begin{eqnarray}
   \eE[\Ni] & \le & \exp_2 \left\{ - N [\Ei(R,p_X,p_{Y|X},F) - \epsilon] \right\} ,
																		\label{eq:Eu} \\
   Pr[\calE_\emptyset] & \le & \exp_2 \left\{
				- N [E_\emptyset (R,p_X,p_{Y|X},F) - \epsilon] \right\} ,
							\quad \forall p_{Y|X} \in \scrW	,			\label{eq:Ee}
\end{eqnarray}
for any $\epsilon > 0$ and $N$ greater than some $N_0(\epsilon)$.
The worst-case exponents (over all $p_{Y|X} \in \scrW$) are denoted by
\begin{eqnarray}
   \Ei(R,p_X,\scrW,F) & \triangleq & \min_{p_{Y|X} \in \scrW} \Ei(R,p_X,p_{Y|X},F) ,
														\label{eq:Eu-W} \\
   E_\emptyset (R,p_X,\scrW,F)
	& \triangleq & \min_{p_{Y|X} \in \scrW} E_\emptyset(R,p_X,p_{Y|X},F) .
														\label{eq:Ee-W}
\end{eqnarray}

Our problem is to maximize the erasure exponent
$E_\emptyset (R,p_X,\scrW,F)$ subject to the constraint that the incorrect-message exponent
$\Ei(R,p_X,\scrW,F)$ is at least equal to some prescribed value $\alpha$.
This is an asymptotic Neyman-Pearson problem.
We shall focus on the regime of practical interest where erasures are more
acceptable than undetected errors: 
\[ E_\emptyset (R,p_X,\scrW,F) \le \Ei(R,p_X,\scrW,F) . \]
We emphasize that asymptotic Neyman-Pearson optimality of the decision rule holds
only in a restricted sense, namely, with respect to the $\calF$-MMI class (\ref{eq:decision}).

Specifically, given $R$ and $\scrW$, we seek the solution to the constrained optimization problem
\begin{equation}
   E_\emptyset^*(R,\scrW,\alpha) \triangleq \max_{p_X} \,\max_{F \in \calF(R,p_X,\scrW,\alpha)}
	\,\min_{p_{Y|X} \in \scrW} E_\emptyset(R,p_X,p_{Y|X},F)
\label{eq:E*}
\end{equation}
where $\calF(R,p_X,\scrW,\alpha)$ is the set of functions $F$ that satisfy
\begin{equation}
   \min_{p_{Y|X} \in \scrW} \Ei(R,p_X,p_{Y|X},F) \ge \alpha
\label{eq:feasible-F0}
\end{equation}
as well as the continuity and monotonicity conditions mentioned above (\ref{eq:decision}).

If we were able to choose $F$ as a function of $p_{Y|X}$, we would do at least
as well as in (\ref{eq:E*}) and achieve the erasure exponent
\begin{eqnarray}
    E_\emptyset^{**}(R,\scrW,\alpha)
	& \triangleq & \max_{p_X} \,\min_{p_{Y|X} \in \scrW} \,\max_{F \in \calF(R,p_X,p_{Y|X},\alpha)}
		E_\emptyset(R,p_X,p_{Y|X},F) \nonumber \\
	& \ge & E_\emptyset^*(R,\scrW,\alpha) .
\label{eq:E**}
\end{eqnarray}
We shall be particularly interested in characterizing $(R,\scrW,\alpha)$ for which 
the decoder incurs no penalty for not knowing $p_{Y|X}$, i.e.,
\begin{itemize}
\item equality holds in (\ref{eq:E**}), and
\item the optimal exponents $\Ei(R,p_X,p_{Y|X},F)$ and $E_\emptyset(R,p_X,p_{Y|X},F)$
	in (\ref{eq:feasible-F0}) and (\ref{eq:E*}) coincide with Forney's exponents
	in (\ref{eq:E-forney}) for all $p_{Y|X} \in \scrW$,
\end{itemize}
the second property being stronger than the first one.

\subsection{Basic Properties of $F$}

To simplify the derivations, it is convenient to slightly strengthen
the requirement that $F$ be nondecreasing, and work with strictly increasing functions $F$
instead. Then the maxima over $F$ in (\ref{eq:E*}) and (\ref{eq:E**}) are replaced
with suprema, but of course their value remains the same.

To each monotonically increasing function $F$ corresponds an inverse $F^{-1}$, such that
\[ F(t) = u \quad \Leftrightarrow \quad F^{-1}(u)=t . \]
Elementary properties satisfied by the inverse function include:
\begin{description}
\item[(P1)] $F^{-1}$ is continuous and increasing over its range.
\item[(P2)] If $F \preceq G$, then $F^{-1} \succeq G^{-1}$.
\item[(P3)] $G(t) = F(t)+\Delta \quad \Leftrightarrow \quad G^{-1}(t) = F^{-1}(t-\Delta)$.
\item[(P4)] $\frac{dF(t)}{dt} = 1 \left/ \frac{dF^{-1}(t)}{dt} \right.$.
\item[(P5)] If $F$ is convex, then $F^{-1}$ is concave.
\item[(P6)] The domain of $F^{-1}$ is the range of $F$, and vice-versa.
\end{description}


Now for any $F$ such that $F(H(p_X)-R) \ge 0$, define the scalar
\begin{equation}
   t_F \triangleq \sup \,\{ t ~:~ F(t) = |F(t')| \quad \forall t' \le t \le H(p_X)-R \}
\label{eq:tF}
\end{equation}
which may depend on $R$ and $p_X$ via the difference $H(p_X)-R$.
From this definition we have the following properties:
\begin{itemize}
\item $|F(t)|^+$ is constant for all $t \le t_F$;
\item $F(t) \ge 0$ for $t \ge t_F$.
\end{itemize}
We have $t_F = 0$ if $F(t)$ is chosen as in (\ref{eq:FR}), or
if $F(t) = \Delta + \lambda |t|^+$.
If $F$ has a zero-crossing, $t_F$ is that zero-crossing. For instance,
$t_F = -\Delta/\lambda$ if $F(t) = \Delta + \lambda t$.
Or $t_F = \min\{ \Delta, H(p_X)-R \}$ if $F(t) = a |t-\Delta|^+$.
The supremum in (\ref{eq:tF}) always exists.


\Section{Random-Coding and Sphere-Packing Exponents}
\label{sec:exp-SingleUser}
\setcounter{equation}{0}

The sphere packing exponent for channel $p_{Y|X}$ is defined as
\begin{equation}
   E_{sp}(R,p_X,p_{Y|X}) \triangleq  \min_{\tp_{Y|X} \,:\,I(p_X, \tp_{Y|X}) \le R} 
	D(\tp_{Y|X} \| p_{Y|X} \,|\,p_X)
\label{eq:Esp}
\end{equation}
and as $\infty$ if the minimization above is over an empty set.
The function $E_{sp}(R,p_X,p_{Y|X})$ is convex, nonincreasing, and differentiable in $R$,
and continuous in $p_{Y|X}$.

The sphere packing exponent for class $\scrW$ is defined as
\begin{equation}
   E_{sp}(R,p_X,\scrW) \triangleq  \min_{p_{Y|X} \in \scrW} E_{sp}(R,p_X,p_{Y|X}) .
\label{eq:Esp-W}
\end{equation}
The function $E_{sp}(R,p_X,\scrW)$ is differentiable in $R$ because $\scrW$ is a connected set.
In some cases, $E_{sp}(R,p_X,\scrW)$ is convex in $R$, e.g., when the same $p_{Y|X}$
achieves the minimum in (\ref{eq:Esp-W}) at all rates.
Denote by $R_\infty(p_X,\scrW)$ the infimum of the rates $R$ such that $E_{sp}(R,p_X,\scrW) < \infty$,
and by $I(p_X,\scrW) = \min_{p_{Y|X} \in \scrW} I(p_X, \,p_{Y|X})$ the supremum of $R$
such that $E_{sp}(R,p_X,\scrW) > 0$.

The {\em modified random coding exponent} for channel $p_{Y|X}$ and for class $\scrW$
are respectively defined as
\begin{equation}
   E_{r,F}(R,p_X,p_{Y|X}) \triangleq \min_{\tp_{Y|X} }
		[D(\tp_{Y|X} \| p_{Y|X} \,|\,p_X) + F(I(p_X, \tp_{Y|X}) - R)]
\label{eq:ErF}
\end{equation}
and
\begin{equation}
   E_{r,F}(R,p_X,\scrW) \triangleq  \min_{p_{Y|X} \in \scrW} E_{r,F}(R,p_X,p_{Y|X}) .
\label{eq:ErF-W}
\end{equation}
When $F(t)=|t|^+$, (\ref{eq:ErF}) is just the usual random coding exponent.
It can be verified that (\ref{eq:ErF}) is a continuous functional of $F$.

\begin{figure}[hbt]
\begin{center}
\includegraphics[width=13cm]{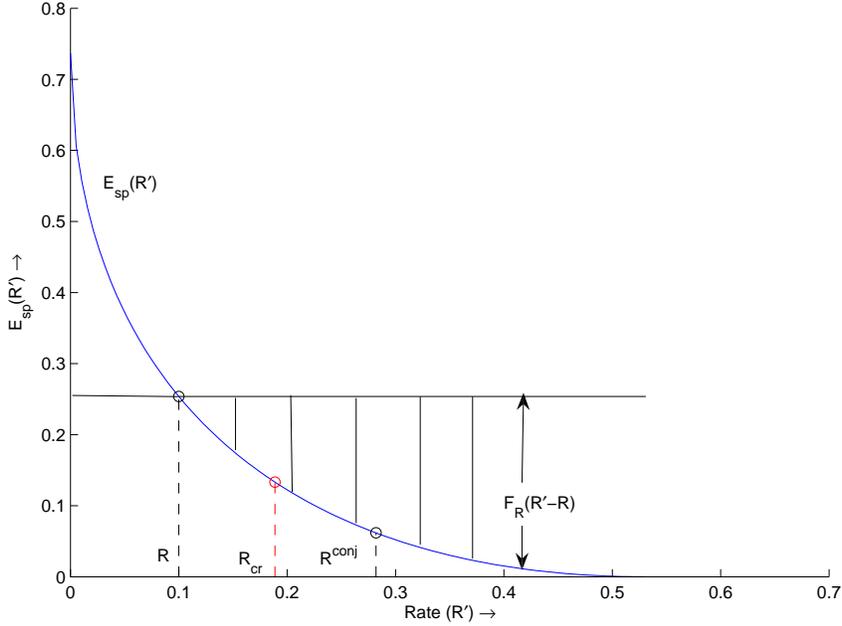}
\end{center}
\caption{Function $F_{R,p_X,\scrW}(\cdot - R)$ when $R=0.1$,
	$\scrW$ is the family of BSC's with crossover probability $\rho \le 0.1$
	(capacity $C(\rho) \ge 0.53$), and $p_X$ is the uniform p.m.f. over $\{0,1\}$.}
\label{fig:FR}
\end{figure}

Define the function
\begin{equation}
   F_{R,p_X,\scrW}(t) \triangleq E_{sp}(R,p_X,\scrW) - E_{sp}(R+t,p_X,\scrW) 
\label{eq:FR}
\end{equation}
which is depicted in Fig.~\ref{fig:FR} for a BSC example to be analyzed in Sec.~\ref{sec:BSC}.
This function is increasing for $R_\infty(p_X,\scrW)-R \le t \le I(p_X,\scrW)-R$
and satisfies the following properties:
\begin{eqnarray}
   F_{R,p_X,\scrW}(0) & = & 0 \nonumber \\
   F_{R,p_X,\scrW}'(t) & = & - E_{sp}'(R+t,p_X,\scrW) \nonumber \\
   E_{sp}(R',p_X,\scrW) + F_{R,p_X,\scrW}(R'-R) & \equiv & E_{sp}(R,p_X,\scrW) .
\label{eq:FR-property}
\end{eqnarray}
If $E_{sp}(R,p_X,\scrW)$ is convex in $R$, then $F_{R,p_X,\scrW}(t)$ is concave in $t$.

\begin{proposition}
The modified random coding exponent $E_{r,F}(R,p_X,p_{Y|X})$ satisfies the following
properties.\\
\noindent
(i) $E_{r,F}(R,p_X,p_{Y|X})$ is nonincreasing in $R$.\\
\noindent
(ii) If $F \prec G$, then $E_{r,F}(R,p_X,p_{Y|X}) \le E_{r,G}(R,p_X,p_{Y|X})$.\\
\noindent
(iii) $E_{r,F}(R,p_X,p_{Y|X})$ is related to the sphere packing exponent as follows:
\begin{equation}
   E_{r,F}(R,p_X,p_{Y|X}) = \min_{R'} [E_{sp}(R',p_X,p_{Y|X}) + F(R'-R)] .
\label{eq:Esp-ErF-1}
\end{equation}
(iv) The above properties hold with $\scrW$ in place of $p_{Y|X}$ in the arguments
of the functions $E_{r,F}$ and $E_{sp}$.
%
\label{prop:ErF-properties}
\end{proposition}

The proof of these properties is given in the appendix. Part (iii) is a variation on
Lemma~5.4 and its corollary in \cite[p.~168]{Csiszar81}.
Also note that while $E_{r,F}(R,p_X,p_{Y|X})$ is convex in $R$
for some choices of $F$, including (\ref{eq:F-CK}), that property
does not extend to arbitrary $F$.

\begin{proposition}
The incorrect-message and erasure exponents under 
the decision rule (\ref{eq:decision}) are respectively given by
\begin{eqnarray}
   \Ei(R,p_X,p_{Y|X},F) & = & E_{r,F}(R,p_X,p_{Y|X}) ,					\label{eq:Eu-F} \\
   E_\emptyset (R,p_X,p_{Y|X},F) & = & E_{r,|F^{-1}|^+}(R,p_X,p_{Y|X}) .\label{eq:Ee-F}
\end{eqnarray}
\label{prop:EuEe}
\end{proposition}
{\em Proof}: see appendix.

If $F(t)=|t|^+$, then $|F^{-1}(t)|^+ = |t|^+$,
and both (\ref{eq:Eu-F}) and (\ref{eq:Ee-F}) reduce to the ordinary
random-coding exponent.
 
If the channel is not reliable enough in the sense that $I(p_X, p_{Y|X}) \le R+F(0)$,
then $F^{-1}(I(p_X, p_{Y|X})-R) \le F^{-1}(F(0)) = 0$, and from (\ref{eq:Ee-F})
we obtain $E_\emptyset (R,p_X,p_{Y|X},F) = 0$ because the minimizing $\tp_{Y|X}$
in the expression for $E_{r,|F^{-1}|^+}$ is equal to $p_{Y|X}$. Checking (\ref{eq:decision}),
a heuristic explanation for the zero erasure exponent is that 
$I(\bx(m);\by) \approx I(p_X,p_{Y|X})$ with high probability when $m$ is the transmitted
message, and $\max_{i \ne m} I(\bx(m);\by) \approx R$
(obtained using (\ref{eq:I-rc}) with $\nu=R$ and the union bound).

\Section{$\scrW$-Optimal Choice of $F$}
\label{sec:optimal-F}
\setcounter{equation}{0}

In this section, we view the incorrect-message and erasure exponents as functionals of $|F|^+$
and examine optimal tradeoffs between them.
It is instructive to first consider the one-parameter family
\begin{equation}
   F(t) \equiv \Delta \ge 0,
\label{eq:F-thr}
\end{equation}
which coresponds to a thresholding rule in (\ref{eq:decision}).
Using (\ref{eq:Esp-ErF-1}) it is easily verified that
\begin{eqnarray}
   \Ei(R,p_X,p_{Y|X},F) = E_{r,F}(R,p_X,p_{Y|X}) & \equiv & \Delta ,      \label{eq:Ei-thr} \\
   E_\emptyset(R,p_X,p_{Y|X},F) & \equiv & E_{sp}(R+\Delta,p_X,p_{Y|X}) . \label{eq:Ee-thr}
\end{eqnarray}
Two extreme choices for $\Delta$ are 0 and $I(p_X,\scrW)-R$ because in each case one error
exponent is zero and the other one is positive. One would expect that better
tradeoffs can be achieved using a broader class of functions $F$, though.

Recalling (\ref{eq:E*}) (\ref{eq:feasible-F0}) and using (\ref{eq:Eu-F}) and (\ref{eq:Ee-F}),
we seek the solution to the following two asymptotic Neyman-Pearson optimization problems.
For list decoding, find
\begin{equation}
   E_\emptyset^L (R,\scrW,\alpha) = \max_{p_X} E_\emptyset^L (R,p_X,\scrW,\alpha)
\label{eq:NP-list-pX}
\end{equation}
where the cost function
\begin{equation}
   E_\emptyset^L (R,p_X,\scrW,\alpha)
	= \sup_{F \in \calF^L(R,p_X,\scrW,\alpha)} E_{r,|F^{-1}|^+}(R,p_X,\scrW) .
\label{eq:NP-list}
\end{equation}
where the feasible set $\calF^L(R,p_X,\scrW,\alpha)$ is the set
of continuous, increasing functions $F$ that yield an incorrect-message exponent
at least equal to $\alpha$:
\begin{equation}
   \Ei(R,p_X,\scrW,F) = E_{r,F}(R,p_X,\scrW) \ge \alpha .
\label{eq:feasible-F}
\end{equation}
For classical decoding (list size $\le 1$), find
\begin{equation}
   E_\emptyset (R,\scrW,\alpha) = \max_{p_X} E_\emptyset (R,p_X,\scrW,\alpha)
\label{eq:NP-pX}
\end{equation}
where
\begin{equation}
   E_\emptyset (R,p_X,\scrW,\alpha)
	= \sup_{F \in \calF(R,p_X,\scrW,\alpha)} E_{r,|F^{-1}|^+}(R,p_X,\scrW)
\label{eq:NP}
\end{equation}
and $\calF(R,p_X,\scrW,\alpha)$ is the subset of functions in
$\calF^L(R,p_X,\scrW,\alpha)$ that satisfy $F(t) \ge t$, i.e., the decoder outputs
at most one message.

Since $\calF(R,p_X,\scrW,\alpha) \subset \calF^L(R,p_X,\scrW,\alpha)$, we have
\[ E_\emptyset (R,p_X,\scrW,\alpha) \le E_\emptyset^L (R,p_X,\scrW,\alpha) . \]
The sequel of this paper focuses on the class of variable-size list decoders
associated with $\calF^L$ because the error exponent tradeoffs are at least
as good as those associated with $\calF$, and the corresponding error exponents
take a more concise form.


Define the critical rate $R_{cr}(p_X,\scrW)$ as the rate at which the derivative
$E_{sp}'(\cdot,p_X,\scrW) = -1$, and
\begin{eqnarray}
	\Delta & \triangleq & \alpha - E_{sp}(R,p_X,\scrW)  . \label{eq:Delta} 
\end{eqnarray}

Two rates $R_1$ and $R_2$ are said to be conjugate given $p_X$ and $\scrW$
if the corresponding slopes of $E_{sp}(\cdot,p_X,\scrW)$ are inverse of each other: 
\begin{equation}
   E_{sp}'(R_1,p_X,\scrW) = \frac{1}{E_{sp}'(R_2,p_X,\scrW)} .
\label{eq:Rconj}
\end{equation}
The difference $d$ between two conjugate rates uniquely specifies them.
We denote the smaller one by $R_1(d)$ and the larger one by $R_2(d)$,
irrespective of the sign of $d$. Hence $R_1(d) \le R_{cr}(p_X,\scrW) \le R_2(d)$,
with equality when $d=0$. We also denote by
$R^{conj}(p_X,\scrW)$ the conjugate rate of $R$, as defined by (\ref{eq:Rconj}).
The conjugate rate always exists when $R$ is {\em below} the critical rate $R_{cr}(p_X,\scrW)$.
If $R$ is above the critical rate and sufficiently large, $R^{conj}(p_X,\scrW)$
may not exist. Instead of treating this case separately, we note that this case
will be irrelevant because the conjugate rate always appears via the expression
$\max\{R, R^{conj}(p_X,\scrW)\}$ which is equal to $R$
if $R > R_{cr}(p_X,\scrW)$ and therefore this expression is always well defined.

The proofs of Prop.~\ref{prop:Fopt} and \ref{prop:Fopt-allR} and Lemma~\ref{lem:E'sup}
below may be found in the appendix; recall $F_{R,p_X,\scrW}(t)$ was defined in (\ref{eq:FR}). 
The proof of Prop.~\ref{prop:Fopt-universal} parallels that of Prop.~\ref{prop:Fopt}(ii)
and is therefore omitted.

\begin{proposition}
The suprema in (\ref{eq:NP-list}) and (\ref{eq:NP}) are respectively achieved by
\begin{eqnarray}
   F^{L*}(t) & = & F_{R,p_X,\scrW}(t) + \Delta 		\label{eq:F-NP-list1} \\
	& = & \alpha - E_{sp}(R+t,p_X,\scrW) 
\label{eq:F-NP-list}
\end{eqnarray}
and
\begin{equation}
   F^*(t) = \max( t, \,F^{L*}(t)) .
\label{eq:F-NP}
\end{equation}
The resulting incorrect-message exponent is given by $\Ei(R,p_X,\scrW) = \alpha$.
The optimal solution is nonunique. In particular, for $t \le 0$, one can
replace $F^{L*}(t)$ by the constant $F^{L*}(0)$ without effect on the error exponents.
\label{prop:Fopt}
\end{proposition}

The proof of Prop.~\ref{prop:Fopt} is quite simple and can be separated from
the calculation of the error exponents. The main idea is as follows.
If $G \preceq F$, we have $E_{r,G}(R,p_X,\scrW) \le E_{r,F}(R,p_X,\scrW)$.
Since $G^{-1} \succeq F^{-1}$, we also have
$E_{r,G^{-1}}(R,p_X,\scrW) \ge E_{r,F^{-1}}(R,p_X,\scrW)$.
Therefore we seek $F^* \in \calF^L(R,p_X,\scrW,\alpha)$ such that $F^* \preceq F$ for all
$F \in \calF^L(R,p_X,\scrW,\alpha)$.
Such $F^*$, assuming it exists, necessarily achieves $E_\emptyset^L (R,p_X,\scrW,\alpha)$.
The same procedure applies to $\calF(R,p_X,\scrW,\alpha)$.

\begin{corollary}
If $R \ge I(p_X,\scrW)$, the thresholding rule $F^{L*}(t) \equiv \Delta$ of (\ref{eq:F-thr})
is optimal, and the optimal error exponents are
$\Ei(R,p_X,\scrW,\alpha)=\Delta$ and $E_\emptyset^L(R,p_X,\scrW,\alpha)=0$.
\label{cor:thr}
\end{corollary}
{\em Proof.}
Since $R \ge I(p_X,\scrW)$, we have $E_{sp}(R,p_X,\scrW) = 0$.
Hence from (\ref{eq:FR}), $F_{R,p_X,\scrW}(t) \equiv 0$
for all $t \ge 0$. Substituting into (\ref{eq:F-NP-list1}) proves the optimality of
the thresholding rule (\ref{eq:F-thr}). The corresponding error exponents are
obtained by minimizing (\ref{eq:Ei-thr}) and (\ref{eq:Ee-thr}) over $p_{Y|X} \in \scrW$.
\hfill $\Box$

The case $R < I(p_X,\scrW)$ is addressed next.

\begin{proposition}
If $E_{sp}(R,p_X,\scrW)$ is convex in $R$,
the optimal incorrect-message and erasure exponents are related as follows. \\
{\bf (i)} For $|R^{conj}(p_X,\scrW)-R|^+ \le \Delta \le I(p_X,\scrW)-R$, we have
\begin{eqnarray}
   \Ei(R,p_X,\scrW,\alpha) & = & E_{sp}(R,p_X,\scrW) + \Delta = \alpha \nonumber \\
	E_\emptyset^L(R,p_X,\scrW,\alpha) & = & E_{sp}(R+\Delta,p_X,\scrW) .
\label{eq:E-regimeI}
\end{eqnarray}
{\bf (ii)} The above exponents are also achieved using the penalty function
$F(t) = \Delta + \lambda |t|^+$ with
\begin{equation}
  - E_{sp}'(R,p_X,\scrW) \le \lambda \le \frac{1}{- E_{sp}'(R+\Delta,p_X,\scrW)} .
\label{eq:lambda-range}
\end{equation}
{\bf (iii)} If $R \le R_{cr}(p_X,\scrW)$ and $0 \le \Delta \le R^{conj}(p_X,\scrW)-R$, we have
\begin{eqnarray}
   \Ei(R,p_X,\scrW,\alpha) & = & E_{sp}(R,p_X,\scrW) + \Delta = \alpha \nonumber \\
	E_\emptyset^L(R,p_X,\scrW,\alpha)
		& = & E_{sp}(R_2(\Delta),p_X,\scrW) + F_{R,p_X,\scrW}^{-1}(R_1(\Delta)-R) .
\label{eq:E-regimeII}
\end{eqnarray}
{\bf (iv)} If $R \le R_{cr}(p_X,\scrW)$ and $R_\infty(p_X, \scrW) - R \le \Delta \le 0$, we have
\begin{eqnarray}
   \Ei(R,p_X,\scrW,\alpha) & = & E_{sp}(R,p_X,\scrW) + \Delta = \alpha \nonumber \\
	E_\emptyset^L(R,p_X,\scrW,\alpha)
		& = & E_{sp}(R_1(\Delta),p_X,\scrW) + F_{R,p_X,\scrW}^{-1}(R_2(\Delta)-R) .
\label{eq:E-regimeIII}
\end{eqnarray}
\label{prop:Fopt-allR}
\end{proposition}

Part (ii) of the proposition implies that not only is the optimal $F$ nonunique
under the combinations of ($R,p_X,\scrW,\alpha$) of Part (i), but also the
Csisz\'{a}r-K\"{o}rner rule (\ref{eq:decision-CK}) is optimal {\em for any}
$(\Delta, \lambda)$ in a certain range of values.

Also, while Prop.~\ref{prop:Fopt-allR} provides simple expressions for
the worst-case error exponents over $\scrW$, the exponents
for any specific channel $p_{Y|X} \in \scrW$ are obtained by substituting the function
(\ref{eq:F-NP-list}) and its inverse, respectively, into the minimization problem of
(\ref{eq:Esp-ErF-1}). This problem does generally not admit a simple expression.

This leads us back to the question asked at the end of Sec.~\ref{sec:decoding}, namely,
when does the decoder pay no penalty for not knowing $p_{Y|X}$?
Defining
\begin{equation}
   \underline{E}_{sp}'(R,p_X,\scrW) \triangleq \min_{p_{Y|X} \in \scrW} E_{sp}'(R,p_X,p_{Y|X}) ,
\label{eq:Esup}
\end{equation}
and
\begin{equation}
   \overline{R}^{conj}(p_X,\scrW) \triangleq \max_{p_{Y|X} \in \scrW} R^{conj}(p_X,p_{Y|X})
\label{eq:Rconj-sup}
\end{equation}
we have the following lemma, whose proof appears in the appendix.
\begin{lemma}
   \begin{eqnarray}
       E_{sp}'(R,p_X,\scrW) & \ge & \underline{E}_{sp}'(R,p_X,\scrW) ,	\label{eq:E'sp-ineq} \\
		\overline{R}^{conj}(p_X,\scrW) & \ge & R^{conj}(p_X,\scrW)		\label{eq:Rconj-ineq}
   \end{eqnarray}
with equality if the same $p_{Y|X}$ minimizes $E_{sp}(R,p_X,p_{Y|X})$ at all rates.
\label{lem:E'sup}
\end{lemma}

\begin{proposition}
Assume that $R$, $p_X$, $\scrW$, $\Delta$ and $\lambda$ are such that
\begin{eqnarray}
   |\overline{R}^{conj}(p_X,\scrW)-R|^+ & \le \Delta & \le I(p_X,\scrW)-R , \label{eq:Delta-range} \\
   - \underline{E}_{sp}'(R,p_X,\scrW) & \le \lambda &
		\le \frac{1}{- \underline{E}_{sp}'(R+\Delta,p_X,\scrW)} .
\label{eq:lambda-range2}
\end{eqnarray}
Then the pair of incorrect-message and erasure exponents 
\begin{eqnarray}
   \{ E_{sp}(R,p_X,p_{Y|X}) + \Delta, \; E_{sp}(R+\Delta,p_X,p_{Y|X}) \}
\label{eq:E-universal}
\end{eqnarray}
is universally attainable over $p_{Y|X} \in \scrW$ using the penalty function
$F(t) = \Delta + \lambda |t|^+$, and equality holds in the erasure-exponent game 
of (\ref{eq:E**}).
\label{prop:Fopt-universal}
\end{proposition}
{\em Proof}.
From (\ref{eq:Rconj-sup}) and (\ref{eq:Delta-range}), we have
\[ |R^{conj}(p_X,p_{Y|X})-R|^+ \le \Delta \le I(p_X,p_{Y|X})-R  , \quad \forall p_{Y|X} \in \scrW . \]
Similarly, from (\ref{eq:Esup}) and (\ref{eq:lambda-range2}), we have
\[ - E_{sp}'(R,p_X,p_{Y|X}) \le \lambda
		\le \frac{1}{- E_{sp}'(R+\Delta,p_X,p_{Y|X})} , \quad \forall p_{Y|X} \in \scrW .
\]
Then applying Prop.~\ref{prop:Fopt-allR}(ii) with the singleton $\{p_{Y|X}\}$ in place of $\scrW$
proves the claim.
\hfill $\Box$

The set of $(\Delta, \lambda)$ defined by (\ref{eq:Delta-range}), (\ref{eq:lambda-range2})
is smaller than that of Prop.~\ref{prop:Fopt-allR}(i) but is not empty because
$\underline{E}_{sp}'(R+\Delta,p_X,\scrW)$ tends to zero as $\Delta$ approaches
the upper limit $I(p_X,\scrW) - R$. Thus the universal exponents in (\ref{eq:E-universal})
hold at least in the small erasure-exponent regime (where $E_{sp}(R+\Delta,p_X,p_{Y|X}) \to 0$)
and coincide with those derived by Forney \cite[Theorem~3(a)]{Forney68} for symmetric
channels, using MAP decoding. For symmetric channels, the same input
distribution $p_X$ is optimal at all rates.
\footnote{
	Forney also studied the case $E_\emptyset(R) > \Ei(R)$, which is not covered
	by our analysis.}
Our rates are identical to his, i.e., the same optimal error exponents are achieved
without knowledge of the channel.


\Section{Relative Minimax}
\label{sec:relative}
\setcounter{equation}{0}

When the compound class $\scrW$ is so large that $I(p_X,\scrW) \le R$, we have
seen from Corollary~\ref{cor:thr} that the simple thresholding rule $F(t) \equiv \Delta$
is optimal. Even if $I(p_X,\scrW) > R$, our minimax criterion (which seeks the worst-case
error exponents over the class $\scrW$) for designing $F$ might be a pessimistic one.
This drawback can be alleviated to some extent using a {\em relative minimax}
principle, see \cite{Feder02} and references therein. Our proposed approach is
to define two functionals
$\alpha(p_{Y|X})$ and $\beta(p_{Y|X})$ and the {\em relative error exponents}
\begin{eqnarray*}
   \Delta_\alpha \,\Ei(R,p_X,p_{Y|X},F) & \triangleq
		& \Ei(R,p_X,p_{Y|X},F) - \alpha(p_{Y|X}) , \\
   \Delta_\beta \,E_\emptyset(R,p_X,p_{Y|X},F) & \triangleq
		& E_{\emptyset}(R,p_X,p_{Y|X},F) - \beta(p_{Y|X}) .
\end{eqnarray*}
Then solve the constrained optimization problem of (\ref{eq:E*}) with the above functionals
in place of $\Ei(R,p_X,p_{Y|X},F)-\alpha$ and $E_{\emptyset}(R,p_X,p_{Y|X},F)$.
It is reasonable to choose $\alpha(p_{Y|X})$ and $\beta(p_{Y|X})$ large for ``good channels''
and small for very noisy channels. While $\alpha(p_{Y|X})$ and $\beta(p_{Y|X})$
could be the error exponents associated with some reference test, this is not a requirement.
A possible choice is
\begin{eqnarray*}
   \alpha(p_{Y|X}) & = & \Delta \\
   \beta(p_{Y|X})  & = & E_{sp}(R+\Delta,p_X,p_{Y|X})
\end{eqnarray*}
which are the error exponents (\ref{eq:Ei-thr}) and (\ref{eq:Ee-thr}) corresponding to
the thresholding rule $F(t) \equiv \Delta$.
Another choice is
\begin{eqnarray}
   \alpha(p_{Y|X}) & = & E_{sp}(R,p_X,p_{Y|X}) + \Delta \label{eq:alpha-relative} \\
   \beta(p_{Y|X})  & = & E_{sp}(R+\Delta,p_X,p_{Y|X})   \label{eq:beta-relative}
\end{eqnarray}
which are the ``ideal'' Forney exponents --- achievable under the assumptions
of Prop.~\ref{prop:Fopt-universal}(i).

The relative minimax problem is a simple extension of the minimax
problem solved earlier. Define the following functions:
\begin{eqnarray}
   \Delta_\alpha \,E_{r,F}(R,p_X,\scrW)
		& = & \min_{p_{Y|X} \in \scrW} [E_{r,F}(R,p_X,p_{Y|X}) - \alpha(p_{Y|X})] , \label{eq:Delta-ErF} \\
   \Delta_\alpha \,E_{sp}(R,p_X,\scrW)
		& = & \min_{p_{Y|X} \in \scrW} [E_{sp}(R,p_X,p_{Y|X}) - \alpha(p_{Y|X})] , \label{eq:Delta-Esp} \\
   F_{R,p_X,\scrW,\alpha}(t) & = & \Delta_\alpha \,E_{sp}(R,p_X,\scrW)
		- \Delta_\alpha \,E_{sp}(R+t,p_X,\scrW) . \label{eq:F-rel}
\end{eqnarray}
The function $F_{R,p_X,\scrW,\alpha}(t)$ of (\ref{eq:F-rel}) is increasing and satisfies
$F_{R,p_X,\scrW,\alpha}(0)=0$. The above functions $\Delta_\alpha \,E_{r,F}$ and 
$\Delta_\alpha \,E_{sp}$ satisfy the following relationship:
\begin{eqnarray}
   \Delta_\alpha \,E_{r,F}(R,p_X,\scrW)
	& \stackrel{(a)}{=} & \min_{p_{Y|X} \in \scrW} \left\{
		\min_{R'} [E_{sp}(R',p_X,p_{Y|X}) + F(R'-R)] - \alpha(p_{Y|X}) \right\}
																	\nonumber \\
	& = & \min_{R'} \left\{ \min_{p_{Y|X} \in \scrW}
		[E_{sp}(R',p_X,p_{Y|X}) - \alpha(p_{Y|X})] + F(R'-R)] \right\} \nonumber \\
	& \stackrel{(b)}{=} & \min_{R'} [\Delta_\alpha \,E_{sp}(R',p_X,\scrW) + F(R'-R)]
\label{eq:Esp-ErF-a}
\end{eqnarray}
where (a) is obtained from (\ref{eq:Esp-ErF-1}) and (\ref{eq:Delta-ErF}),
and (b) from (\ref{eq:Delta-Esp}).
Equation (\ref{eq:Esp-ErF-a}) is of the same form as (\ref{eq:Esp-ErF-1}),
with $\Delta_\alpha \,E_{r,F}$ and $\Delta_\alpha \,E_{sp}$ in place of $E_{r,F}-\alpha$
and $E_{sp}-\alpha$, respectively.

Analogously to (\ref{eq:NP-list-pX}), (\ref{eq:NP-list}), and (\ref{eq:feasible-F}),
the relative minimax for variable-size decoders is given by
\begin{equation}
   \Delta_\beta \,E_\emptyset^L(R,\scrW,\alpha)
	= \max_{p_X} \sup_{F \in \calF^L(R,p_X,\scrW,\alpha)}
		\Delta_\beta \,E_{r,|F^{-1}|^+}(R,p_X,\scrW)
\label{eq:relative}
\end{equation}
where the feasible set $\calF^L(R,p_X,\scrW,\alpha)$ is the set of functions $F$ that satisfy
\[ \Delta_\alpha \,E_{r,F}(R,p_X,\scrW) \ge 0 \]
as well as the previous continuity and monotonicity conditions.
The following proposition is analogous to Prop.~\ref{prop:Fopt}.

\begin{proposition}
The supremum over $F$ in (\ref{eq:relative}) is achieved by
\begin{eqnarray}
   F^{L*}(t) & = & F_{R,p_X,\scrW,\alpha}(t) - \Delta_\alpha \,E_{sp}(R,p_X,\scrW) \nonumber \\
	& = & - \Delta_\alpha \,E_{sp}(R+t,p_X,\scrW) ,
\label{eq:Fopt-relative}
\end{eqnarray}
independently of the choice of $\beta$.
The relative minimax is given by
\[ \Delta_\beta E_\emptyset^L(R,\scrW,\alpha)
	= \max_{p_X} \Delta_\beta \,E_{r,|(F^{L*})^{-1}|}(R,p_X,\scrW) . \]
\end{proposition}
{\em Proof}: The proof exploits the same monotonicity property ($E_{r,F} \le E_{r,G}$
for $F \preceq G$) that was used to derive the optimal $F$ in (\ref{eq:F-NP-list}).
The supremum over $F$ is obtained by following the steps of the proof of Prop.~\ref{prop:Fopt},
substituting $\Delta_\alpha \,E_{r,F}$, $\Delta_\alpha \,E_{sp}$, and $F_{R,p_X,\scrW,\alpha}$
for $E_{r,F}-\alpha$, $E_{sp}-\alpha$, and $F_{R,p_X,\scrW}$, respectively.
The relative minimax is obtained by substituting the optimal $F$ into (\ref{eq:relative}).
\hfill $\Box$

We would like to know how much influence the reference function $\alpha$ has on the optimal $F$.
For the ``Forney reference exponent function'' $\alpha$ of (\ref{eq:alpha-relative}),
we obtain the optimal $F$ from (\ref{eq:Fopt-relative}) and (\ref{eq:Delta-Esp}):
\begin{eqnarray}
   F^{L*}(t)
	& = & - \min_{p_{Y|X} \in \scrW} [E_{sp}(R+t,p_X,p_{Y|X}) - \alpha(p_{Y|X})] \nonumber \\
	& = & - \min_{p_{Y|X} \in \scrW}
					[E_{sp}(R+t,p_X,p_{Y|X}) - E_{sp}(R,p_X,p_{Y|X}) - \Delta] \nonumber \\
	& = & \Delta + \max_{p_{Y|X} \in \scrW}
					[E_{sp}(R,p_X,p_{Y|X}) - E_{sp}(R+t,p_X,p_{Y|X})] \nonumber \\
	& = & \Delta + \max_{p_{Y|X} \in \scrW} F_{R,p_X,p_{Y|X}}(t) .
\label{eq:Fopt-relative-Esp}
\end{eqnarray}
Interestingly, the maximum above is often achieved by the {\em cleanest} channel
in $\scrW$ --- for which $E_{sp}(R,p_X,p_{Y|X})$ is large and $E_{sp}(R+t,p_X,p_{Y|X})$
falls off rapidly as $t$ increases.
This stands in contrast to (\ref{eq:F-NP-list}) which may be written as
\begin{eqnarray}
   F^{L*}(t) & = & \alpha - \min_{p_{Y|X} \in \scrW} E_{sp}(R+t,p_X,p_{Y|X}) \nonumber \\
	& = & \Delta + \min_{p_{Y|X} \in \scrW}  E_{sp}(R,p_X,p_{Y|X})
				- \min_{p_{Y|X} \in \scrW}  E_{sp}(R+t,p_X,p_{Y|X}) .
\label{eq:Fopt-2}
\end{eqnarray}
In (\ref{eq:Fopt-2}), the minima are achieved by the {\em noisiest} channel at rates
$R$ and $R+t$, respectively. Also note that $F^{L*}(t)$ from (\ref{eq:Fopt-relative-Esp}) 
is uniformly larger than $F^{L*}(t)$ from (\ref{eq:Fopt-2}) and thus results in larger
incorrect-message exponents.

For $R > I(p_X,\scrW)$, Corollary~\ref{cor:thr} has shown
that the minimax criterion is maximized by the thresholding rule $F(t)=\Delta$ which
yields $\Ei(R,p_X,p_{Y|X}) = \Delta$ and
$E_\emptyset(R,p_X,p_{Y|X}) = E_{sp}(R+\Delta,p_X,p_{Y|X})$ for all $p_{Y|X}$.
The relative minimax criterion based on $\alpha(p_{Y|X})$ of (\ref{eq:alpha-relative})
yields a higher $\Ei(R,p_X,p_{Y|X})$ for good channels and this is counterbalanced
by a lower $E_\emptyset(R,p_X,p_{Y|X})$. Thus the primary advantage of the relative minimax
approach is that $\alpha(p_{Y|X})$ can be chosen to more finely balance the error
exponents across the range of channels of interest.


\Section{Compound Binary Symmetric Channel}
\label{sec:BSC}
\setcounter{equation}{0}

We have evaluated the incorrect-message and erasure exponents of (\ref{eq:E-universal})
for the compound BSC with crossover probability $\rho \in [\rho_{\min}, \rho_{\max}]$,
where $0 < \rho_{\min} < \rho_{\max} \le \frac{1}{2}$.
The class $\scrW$ may be identified with the interval $[\rho_{\min} ,\, \rho_{\max}]$,
where $\rho_{\min}$ and $\rho_{\max}$ correspond to the cleanest and noisiest channels
in $\scrW$, respectively.
Denote by $h_2(\rho) \triangleq - \rho \log \rho - (1-\rho) \log (1-\rho)$  the binary entropy
function, by $h_2^{-1}(\cdot)$ the inverse of that function over the range $[0, \frac{1}{2}]$,
and by $p_{\rho}$ the Bernoulli p.m.f. with parameter $\rho$.

Capacity of the BSC is given by $C(\rho) = 1 - h_2(\rho)$, and the sphere packing exponent
by \cite[p.~195]{Csiszar81}
\begin{eqnarray}
   E_{sp}(R,\rho)
	& = & D(p_{\rho_R}||p_\rho) \nonumber \\
	& = & \rho_R \log \frac{\rho_R}{\rho} + (1-\rho_R) \log \frac{1-\rho_R}{1-\rho}  ,
	\quad 0 \le R \le C(\rho) ,
\label{eq:Esp-BSC}
\end{eqnarray}
where $\rho_R = h_2^{-1}(1-R) \ge \rho$.
The optimal input distribution $p_X$ is uniform at all rates and will be omitted
from the list of arguments of the functions $E_{r,F}$, $E_{sp}$, and $F_R$ below.
The critical rate is
\[ R_{cr}(\rho) = 1 - h_2 \left( \frac{1}{1+\sqrt{1/\rho^2 - 1}} \right) , \]
and $E_{sp}(0,\rho) = - \log \sqrt{4\rho(1-\rho)}$.

The capacity and sphere-packing exponent for the compound BSC are respectively
given by $C(\scrW) = C(\rho_{\max})$ and
\begin{equation}
   E_{sp}(R,\scrW) = \min_{\rho_{\min} \le \rho \le \rho_{\max}} E_{sp}(R,\rho)
	= E_{sp}(R,\rho_{\max}) .
\label{eq:Esp-min-BSC}
\end{equation}
For $R \ge C(\scrW)$, the optimal $F$ is the thresholding rule of (\ref{eq:F-thr}),
and (\ref{eq:Ei-thr}) and (\ref{eq:Ee-thr}) yield
\[ \Ei(R,\rho)=\Delta \quad \mathrm{and} \quad
	E_\emptyset(R,\rho) = E_{sp}(R+\Delta,\rho) . \]
In the remainder of this section we consider the case $R < C(\scrW)$,
in which case $E_{sp}(R,\scrW) > 0$.

{\bf Optimal $F$.}
For any $0 \le \Delta \le C(\rho)-R$, we have
$\rho \le \rho_{R+\Delta} \le \rho_R \le \frac{1}{2}$.
From (\ref{eq:Esp-BSC}) we have
\begin{eqnarray*}
   F_{R,\rho}(t) & = & E_{sp}(R,\rho) - E_{sp}(R+t,\rho) \\
	& = & D(p_{\rho_R}||p_\rho) - D(p_{\rho_{R+t}}||p_\rho) \\
	& = & h_2(\rho_R) - h_2(\rho_{R+t}) + (\rho_R - \rho_{R+t})
			\left( \log\frac{1}{\rho} - \log \frac{1}{1-\rho} \right) \\
	& = & h_2(\rho_R) - h_2(\rho_{R+t}) + (\rho_R - \rho_{R+t}) 
			\log \left(\frac{1}{\rho} - 1 \right) , \quad t \ge 0 ,
\end{eqnarray*}
which is a decreasing function of $\rho$.

Evaluating the optimal $F$ from (\ref{eq:F-NP-list}), we have
\begin{eqnarray*}
   F^{L*}(t)
	& = & \Delta + F_{R,\scrW}(t) \\
	& = & \Delta + E_{sp}(R,\scrW) - E_{sp}(R+t,\scrW) \\
	& = & \Delta + E_{sp}(R,\rho_{\max}) - E_{sp}(R+t,\rho_{\max}) \\
	& = & \Delta + F_{R,\rho_{\max}}(t) , \quad t \ge 0 .
\end{eqnarray*}
Observe that the optimal $F$  is determined by the noisiest channel ($\rho_{\max}$)
and does not depend at all on $\rho_{\min}$.

This contrasts with the relative minimax criterion with $\alpha(p_{Y|X})$ of
(\ref{eq:alpha-relative}), where evaluation of the optimal $F$ from
(\ref{eq:Fopt-relative-Esp}) yields
\begin{eqnarray*}
   F^{L*}(t)
	& = & \Delta + \max_{\rho_{\min} \le \rho \le \rho_{\max}} F_{R,\rho}(t) \\
	& = & \Delta + F_{R,\rho_{\min}}(t) , \quad t \ge 0
\end{eqnarray*}
which is determined by the cleanest channel ($\rho_{\min}$)
and does not depend on $\rho_{\max}$. 

{\bf Optimal Error Exponents.}
The derivations below are simplified if instead of working with the crossover
probability $\rho$, we use the following reparameterization:
\[ \mu \triangleq \rho^{-1} - 1, \quad \mu_{\max} \triangleq \rho_{\min}^{-1} - 1,
	\quad \mu_{\min} \triangleq \rho_{\max}^{-1} - 1 \]
and
\[ \mu_R \triangleq \rho_R^{-1} - 1 = \frac{1}{h_2^{-1}(1-R)} - 1 
	\quad \Leftrightarrow \quad R = 1 - h_2 \left( \frac{1}{1+\mu} \right) \]
where $\mu_R$ increases monotonically from 1 to $\infty$ as $R$ increases from 0 to 1.
With this notation, we have $\rho = \frac{1}{1+\mu}$ and 
$\mu_{\max} \ge \mu \ge \mu_{R+\Delta} \ge \mu_R \ge 1$. Also
\[ \frac{dR}{d\rho_R} = - \frac{dh_2(\rho_R)}{d\rho_R} = - \frac{\log \mu_R}{\ln 2} \]
\[ \frac{d E_{sp}(R,\rho)}{d\rho_R} = \frac{\log \mu - \log \mu_R}{\ln 2} \]
\begin{equation}
   - E_{sp}'(R,\rho) = \frac{d E_{sp}(R,\rho)/d\rho_R}{-dR/d\rho_R}
	= \frac{\log \mu}{\log \mu_R} - 1 = \frac{\log \mu/\mu_R}{\log \mu_R} .
\label{eq:Esp'-BSC}
\end{equation}
From (\ref{eq:Esup}) and (\ref{eq:Esp'-BSC}), we obtain
\begin{equation}
   - \underline{E}_{sp}'(R,\scrW)
	= - \min_{\rho_{\min} \le \rho \le \rho_{\max}} E_{sp}'(R,\rho)
	= \frac{\log \mu_{\max}/\mu_R}{\log \mu_R}  .
\label{eq:min-Esp'-BSC}
\end{equation}
Observe that the minimizing $\rho$ is $\rho_{\min}$, i.e., the cleanest channel in $\scrW$.
In contrast, from (\ref{eq:Esp-min-BSC}), we have
\begin{equation}
   - E_{sp}'(R,\scrW)
	= - E_{sp}'(R,\rho_{\max}) = \frac{\log \mu_{\min}/\mu_R}{\log \mu_R} 
	< - \underline{E}_{sp}'(R,\scrW)
\label{eq:Esp'-W-BSC}
\end{equation}
which is determined by the noisiest channel ($\rho_{\max}$).

{\bf Conditions for universality.}
Next we evaluate $\overline{R}^{conj}(\scrW)$ from (\ref{eq:Rconj-sup}).
For a given $\rho$, the conjugate rate of $R$ is obtained from (\ref{eq:Esp'-BSC}):
\begin{eqnarray*}
	- E_{sp}'(R,\rho) & = & \frac{1}{- E_{sp}'(R^{conj},\rho)} \\
	\frac{\log(\mu/\mu_R)}{\log \mu_R} & = & \frac{\log \mu_{R^{conj}}}{\log (\mu/\mu_{R^{conj}})}
\end{eqnarray*}
hence
\begin{eqnarray*}
   \mu_{R^{conj}} & = & \frac{\mu}{\mu_R} \\
    R^{conj}(\mu) & = & 1 - h_2 \left( \frac{1}{1+\mu/\mu_R} \right) \\
   \overline{R}^{conj}(\scrW) & = & \max_{\mu_{\min} \le \mu \le \mu_{\max}} R^{conj}(\mu) \\
		& = & R^{conj}(\mu_{\max}) . 
\end{eqnarray*}
From (\ref{eq:Rconj}) and (\ref{eq:Esp-min-BSC}), we have
\[ R^{conj}(\scrW) = R^{conj}(\mu_{\min}) . \]
Analogously to (\ref{eq:Esp'-W-BSC}), observe that both $\overline{R}^{conj}(\scrW)$ and
$R^{conj}(\scrW)$ are determined by the cleanest and noisiest channels in $\scrW$,
respectively.

We can now evaluate the conditions of Prop.~\ref{prop:Fopt-universal}, under which
$F(t)=\Delta + \lambda |t|^+$ is universal (subject to conditions on $\Delta$ and $\lambda$).
In (\ref{eq:Delta-range}), $\Delta$ must satisfy
\begin{eqnarray}
   \left| \overline{R}^{conj}(\scrW)-R \right|^+ \le & \Delta & \le C(\scrW) - R \nonumber \\
  \left| R^{conj}(\mu_{\max})-R \right|^+ \le & \Delta & \le C(\mu_{\min}) - R \nonumber \\
   \left| h_2 \left( \frac{1}{1+\mu_R} \right) - h_2 \left( \frac{1}{1+\mu_{\max}/\mu_R} \right)
	\right|^+ \le & \Delta & \le h_2 \left( \frac{1}{1+\mu_R} \right)
		- h_2 \left( \frac{1}{1+\mu_{\min}} \right) .
\label{eq:Delta-range-BSC}
\end{eqnarray}
The left side is zero if $\mu_{\max} \le \mu_R^2$.
If $\mu_{\max} > \mu_R^2$, the argument of $|\cdot|^+$ is positive, and we need
$\mu_{\max}/\mu_R < \mu_{\min}$ to ensure that the left side is lower than the right side.
Hence there exists a nonempty range of values of $\Delta$ satisfying (\ref{eq:Delta-range-BSC})
if and only if
\[ \mu_R \ge \min \left\{ \sqrt{\mu_{\max}} , \, \frac{\mu_{\max}}{\mu_{\min}} \right\} \]
which may also be written as
\[ \mu_{\max} \le \max \{ \mu_R^2 , \,\mu_R \,\mu_{\min} \} . \]

Next, substituting (\ref{eq:min-Esp'-BSC}) into (\ref{eq:lambda-range2}), we obtain
the following condition for $\lambda$:
\begin{equation}
    \frac{\log \mu_{\max}/\mu_R}{\log \mu_R} \le \lambda
	\le \frac{\log \mu_{R+\Delta}}{\log \mu_{\max}/\mu_{R+\Delta}} .
\label{eq:lambda-rho}
\end{equation}
This equation has a solution if and only if the left side does not exceed the right side, i.e.,
\begin{equation}
   \mu_{\max} \le \mu_R \,\mu_{R+\Delta} ,
\label{eq:mu-max-ineq}
\end{equation}
or equivalently, $\rho_{\min} \ge (1+ \mu_R \,\mu_{R+\Delta})^{-1}$.
Since $\mu_R$ is an increasing function of $R$, the larger the values of $R$ and $\Delta$,
the lower the value of $\rho_{\min}$ for which the universality property still holds. 

If equality holds in (\ref{eq:mu-max-ineq}), the only feasible value of $\lambda$ is
\begin{equation}
   \lambda = \frac{\log \mu_{R+\Delta}}{\log \mu_R} \ge 1 .
\label{eq:lambda-opt}
\end{equation}
This value of $\lambda$ remains feasible if (\ref{eq:mu-max-ineq}) holds
with strict inequality.

\begin{figure}[h!]
\begin{center}
\includegraphics[width=13cm]{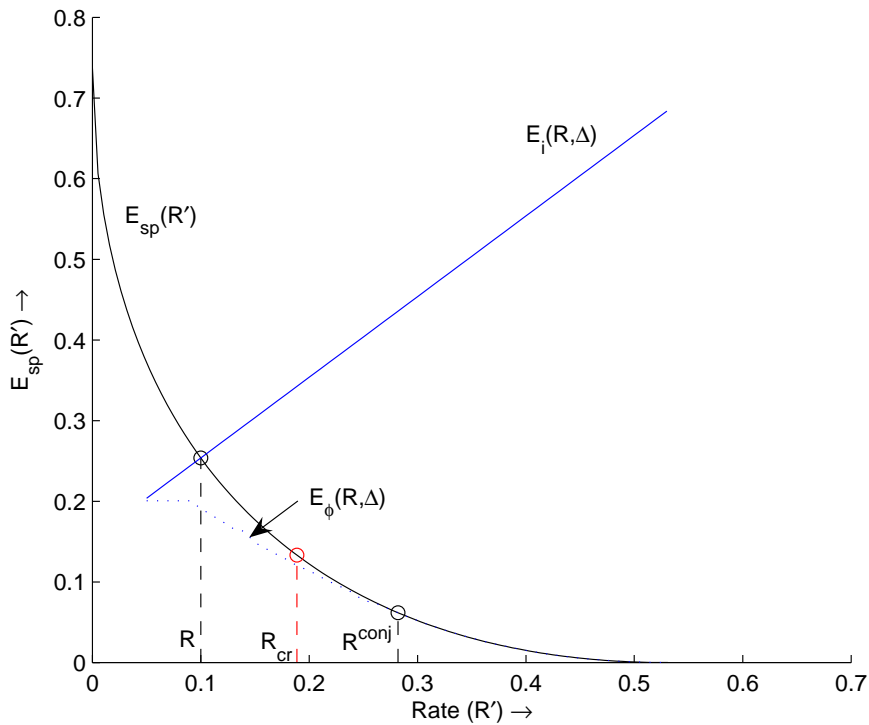}
\end{center}
\caption{Erasure and incorrect-message exponents $E_{\emptyset}(R)$ and $\Ei(R)$
for BSC with crossover probability $\rho=0.1$ (capacity $C(\rho) \approx 0.53$)
and rate $R=0.1$.}
\label{fig:BSC}
\end{figure}

\Section{Discussion}
\label{sec:discussion}
\setcounter{equation}{0}

The $\calF$-MMI decision rule of (\ref{eq:decision}) is a generalization of
Csisz\'{a}r and K\"{o}rner's MMI decoder with erasure option.
The weighting function $F$ in (\ref{eq:decision}) can be optimized
in an asymptotic Neyman-Pearson sense given a compound class of channels $\scrW$.
An explicit formula has been derived in terms of the sphere-packing exponent
function for $F$ that maximizes the erasure exponent subject to a constraint
on the incorrect-message exponent. The optimal $F$ is generally nonunique
but agrees with existing designs in special cases of interest.

In particular, Corollary~\ref{cor:thr} shows that the simple thresholding rule
$F(t) \equiv \Delta$ is optimal if $R \ge I(p_X,\scrW)$, i.e., when the transmission
rate cannot be reliably supported by the worst channel in $\scrW$.
When $R < I(p_X,\scrW)$, Prop.~\ref{prop:Fopt-universal} shows that for small erasure
exponents, our expressions for the optimal exponents coincide with those
derived by Forney \cite{Forney68} for symmetric channels, where the same input
distribution $p_X$ is optimal at all rates. In this regime,
Csisz\'{a}r and K\"{o}rner's rule $F(t) = \Delta + \lambda |t|^+$ is also universal
under some conditions on the parameter pair $(\Delta, \lambda)$.
It is also worth noting that while suboptimal, the design $F(t) = \Delta + t$ yields
an empirical version of Forney's simple decision rule (\ref{eq:decision-forney-2}).

Previous work \cite{Merhav07} using a different universal decoder 
had shown that Forney's exponents can be matched in the special case where
undetected-error and erasure exponents are equal
(corresponding to $T=0$ in Forney's rule (\ref{eq:decision-forney})).
Our results show that this property extends beyond this special case,
albeit not everywhere.

Another analogy between Forney's suboptimal decision rule (\ref{eq:decision-forney-2}) 
and ours (\ref{eq:decision}) is that the former is based on the two highest likelihood
scores, and the latter is based on the two highest empirical mutual information scores.
Our results imply that (\ref{eq:decision-forney-2}) in optimal
(in terms of error exponents) in the special regime identified above.

The relative minimax criterion of Sec.~\ref{sec:relative} is attractive
when the compound class $\scrW$ is broad (or difficult to pick) as it allows finer tuning
of the error exponents for different channels in $\scrW$. The class $\scrW$ could
conceivably be as large as $\scrP_{Y|X}$, the set of all DMC's.

Finally, we have extended our framework to decoding for compound
multiple access channels. Those results will be presented elsewhere.

{\bf Acknowledgements}. The author thanks Shankar Sadasivam for numerical
evaluation of the error exponent formulas in Sec.~\ref{sec:BSC},
and Prof. Lizhong Zheng for helpful comments.

\newpage

\appendix

\renewcommand{\theequation}{\Alph{section}.\arabic{equation}}

\Section{Proof of Proposition~\ref{prop:ErF-properties}}

\noindent
(i) is immediate from (\ref{eq:ErF}), restated below:
\begin{equation}
   E_{r,F}(R,p_X,p_{Y|X}) = \min_{\tp_{Y|X} }
		[D(\tp_{Y|X} \| p_{Y|X} \,|\,p_X) + F(I(p_X, \tp_{Y|X}) - R) ] ,
\label{eq:ErF-pYX}
\end{equation}
and the fact that $F$ is nondecreasing.\\
(ii) is immediate for the same reason as above.\\
(iii) Since the function $E_{sp}(R,p_X,p_{Y|X})$ is decreasing in $R$ 
for $R \le I(p_X, p_{Y|X})$, we have
\begin{equation}
   E_{sp}(R,p_X,p_{Y|X}) = \min_{\tp_{Y|X} \,:\,I(p_X, \tp_{Y|X}) = R} 
			D(\tp_{Y|X} \| p_{Y|X} \,|\,p_X) , \quad \forall R \le I(p_X, p_{Y|X}) .
\label{eq:Esp=}
\end{equation}

Since $F$ is nondecreasing, $\tp_{Y|X}$ that achieves the minimum in (\ref{eq:ErF-pYX})
must satisfy $I(p_X, \tp_{Y|X}) \le I(p_X, p_{Y|X})$.
Hence
\begin{eqnarray*}
   \lefteqn{E_{r,F}(R,p_X,p_{Y|X})} \\
	& = & \min_{\tp_{Y|X} \,:\, I(p_X, \tp_{Y|X}) \le I(p_X, p_{Y|X})}
			[D(\tp_{Y|X} \| p_{Y|X} \,|\,p_X) + F(I(p_X, \tp_{Y|X})-R)] \\
	& = & \min_{R' \le I(p_X, p_{Y|X})}
			\,\min_{\tp_{Y|X} \,:\,I(p_X, \tp_{Y|X}) = R'}
			[D(\tp_{Y|X} \| p_{Y|X} \,|\,p_X) + F(R'-R)] \\
	& \stackrel{(a)}{=} & \min_{R' \le I(p_X, p_{Y|X})}
			[E_{sp}(R',p_X,p_{Y|X})+ F(R'-R)] \\
	& \stackrel{(b)}{=} & \min_{R'} [E_{sp}(R',p_X,p_{Y|X})+ F(R'-R)]
\end{eqnarray*}
where (a) is due to (\ref{eq:Esp=}), and
(b) holds because $E_{sp}(R',p_X,p_{Y|X}) = 0$ for $R' \ge I(p_X, p_{Y|X})$
and $F$ is nondecreasing. \\
(iv) The claim follows directly from the definitions (\ref{eq:Esp-W}) and (\ref{eq:ErF-W}),
taking minima over $\scrW$.
%
%
\hfill $\Box$

\Section{Proof of Proposition~\ref{prop:EuEe}}

Given the p.m.f. $p_X$, choose any type $p_{\bx}$ such that
$\max_{x \in \calX} |p_{\bx}(x) - p_X(x)| \le \frac{|\calX|}{N}$. \footnote{
   For instance, truncate each $p_X(x)$ down to the nearest integer multiple of $1/N$
   and add $a/N$ to the smallest resulting value to obtain $p_{\bx}(x), x \in \calX$,
   summing to one. $a$ is an integer in the range $\{ 0, 1, \cdots, |\calX|-1 \}$.
   }
Define
\begin{equation}
   E_{r,F,N}(R,p_{\bx},p_{Y|X}) = \min_{p_{\by|\bx}} 
	\,[D(p_{\by|\bx} \| p_{Y|X} | p_{\bx}) + F(I(\bx;\by) - R)]
\label{eq:ErFN}
\end{equation}
and
\begin{equation}
   E_{sp,N}(R,p_{\bx},p_{Y|X}) = \min_{p_{\by|\bx} \,:\, I(\bx;\by) \le R}
				D(p_{\by|\bx} \| p_{Y|X} | \,p_{\bx})
\label{eq:EspN}
\end{equation}
which differ from (\ref{eq:ErF}) and (\ref{eq:Esp}) in that the minimization
is performed over conditional types instead of general conditional p.m.f.'s. We have
\begin{eqnarray}
   \lim_{N \to \infty} E_{r,F,N}(R,p_{\bx},p_{Y|X}) & = & E_{r,F}(R,p_X,p_{Y|X}) \nonumber \\
	\lim_{N \to \infty} E_{sp,N}(R,p_{\bx},p_{Y|X}) & = & E_{sp}(R,p_X,p_{Y|X})
\label{eq:lim-EN}
\end{eqnarray}
by continuity of the divergence functional and of $F$ in the region
where the minimand $D+F$ is finite.

We will use the following two standard inequalities.


1) Given an arbitrary sequence $\by$, draw $\bx'$ independently of $\by$ and
uniformly over a fixed type class $T_{\bx}$. Then \cite{Csiszar81}
\[ Pr[T_{\bx'|\by}] = \frac{|T_{\bx'|\by}|}{|T_{\bx}|}
	= \frac{|T_{\bx'|\by}|}{|T_{\bx'}|} \doteq 2^{-N I(\bx';\by)} . \]
Hence for any $0 \le \nu \le H(p_{\bx})$,
\begin{eqnarray}
   Pr[I(\bx';\by) \ge \nu]
	& = & \sum_{T_{\bx'|\by}} Pr[T_{\bx'|\by}] \,\mathds1_{\{I(\bx';\by) \ge \nu\} } \nonumber \\
	& \doteq & \sum_{T_{\bx'|\by}} 2^{-N I(\bx';\by)} \,\mathds1_{\{I(\bx';\by) \ge \nu\} } 
																	\nonumber \\
	& \stackrel{(a)}{\doteq} & \max_{T_{\bx'|\by}} 2^{-N I(\bx';\by)}
			\,\mathds1_{\{I(\bx';\by) \ge \nu\} } 					\nonumber \\
	& \doteq & 2^{-N\nu}
\label{eq:I-rc}
\end{eqnarray}
where (a) holds because the number of types is polynomial in $N$.
For $\nu > H(p_{\bx})$ we have $Pr[I(\bx';\by) \ge \nu] = 0$.

2) Given an arbitrary sequence $\bx$, draw $\by$ from the conditional p.m.f. $p_{Y|X}^N(\cdot|\bx)$.
We have \cite{Csiszar81}
\begin{equation}
   Pr[T_{\by|\bx}] \doteq 2^{-N D(p_{\by|\bx} \| p_{Y|X} | p_{\bx})} .
\label{eq:PTyx}
\end{equation}
Then for any $\nu > 0$,
\begin{eqnarray}
   Pr[I(\bx;\by) \le \nu] 
	& = & \sum_{T_{\by|\bx}} Pr[T_{\by|\bx}] \,\mathds1_{\{I(\bx;\by) \le \nu\} } \nonumber \\
	& \doteq & \sum_{p_{\by|\bx}} 2^{-N D(p_{\by|\bx} \| p_{Y|X} | p_{\bx})}
				\,\mathds1_{\{I(\bx;\by) \le \nu\} } \nonumber \\
	& \doteq & \max_{p_{\by|\bx}} 2^{-N D(p_{\by|\bx} \| p_{Y|X} | p_{\bx})}
				\,\mathds1_{\{I(\bx;\by) \le \nu\} } \nonumber \\
	& = & \max_{p_{\by|\bx}~:~I(\bx;\by) \le \nu} 2^{-N D(p_{\by|\bx} \| p_{Y|X} | p_{\bx})} 
																			\nonumber \\
	& = & 2^{-N E_{sp,N}(\nu,p_{\bx},p_{Y|X})} .
\label{eq:I-sp}
\end{eqnarray}


{\bf Incorrect Messages.}
The codewords are drawn independently and uniformly from type class $T_{\bx}$.
Since the conditional error probability is independent of the transmitted message,
assume without loss of generality that message $m=1$ was transmitted.
An incorrect codeword $\bx(i)$ appears on the decoder's list if $i > 1$ and
\[ I(\bx(i);\by) \ge R + \max_{j \ne i} F(I(\bx(j);\by) - R) . \]
Let $\bx = \bx(1)$.
To evaluate the expected number of incorrect codewords on the list,
we first fix $\by$.

Given $\by$, define the i.i.d. random variables $Z_i = I(\bx(i);\by)-R$ for $2 \le i \le 2^{NR}$.
Also let $z_1 = I(\bx;\by)-R$, which is a function of the joint type $p_{\bx\by}$.
The expected number of incorrect codewords on the list depends on ($\bx,\by$)
only via their joint type and is given by
\begin{eqnarray}
  \eE[\Ni|T_{\bx\by}]
	& = & \sum_{i=2}^{2^{NR}} \,Pr \left[ I(\bx(i);\by) \ge R + \max_{j \notin \calM\setminus\{i\}}
												F(I(\bx(j);\by) - R) \right] \nonumber \\
    & = & \sum_{i=2}^{2^{NR}} \,Pr \left[ Z_i \ge \max_{j \notin \calM\setminus\{i\}}
												F(Z_j) \right] \nonumber \\
	& \le & \sum_{i=2}^{2^{NR}} \,Pr[Z_i \ge F(z_1)] \nonumber \\
	& = & (2^{NR} -1) \,Pr[Z_2 \ge F(z_1)] \nonumber \\
	& \stackrel{(a)}{=} & (2^{NR} -1) \,Pr[I(\bx';\by) \ge R + F(I(\bx;\by) - R)] \nonumber \\
	& \stackrel{(b)}{\doteq} & 2^{NR} \,2^{-N[R + F(I(\bx;\by) - R)]} \nonumber \\
	& = & 2^{-N F(I(\bx;\by) - R)}
\label{eq:ENi-ineq}
\end{eqnarray}
where in (a), $\bx'$ is drawn independently of $\by$ and uniformly over the type class $T_\bx$;
and (b) is obtained by application of (\ref{eq:I-rc}).

Averaging over $\by$, we obtain
\begin{eqnarray*}
   \eE[\Ni|T_{\bx}]
	& = & \sum_{T_{\by|\bx}} Pr[T_{\by|\bx}] \,\eE[\Ni|T_{\bx\by}] \\
	& \doteq & \max_{T_{\by|\bx}} Pr[T_{\by|\bx}] \,\eE[\Ni|T_{\bx\by}] \\
	& \stackrel{(a)}{\doteq} & \max_{p_{\by|\bx}} \exp_2 \{ -N [D(p_{\by|\bx} \| p_{Y|X} | p_{\bx}) 
					+ F(I(\bx;\by) - R)] \} \\
	& \stackrel{(b)}{=} & \exp_2 \{ -N E_{r,F,N}(R,p_{\bx},p_{Y|X}) \} \\
	& \stackrel{(c)}{\doteq} & \exp_2 \{ -N E_{r,F}(R,p_X,p_{Y|X}) \} 
\end{eqnarray*}
where (a), (b), (c) follow from (\ref{eq:PTyx}), (\ref{eq:ErFN}), (\ref{eq:lim-EN}), respectively.
This proves (\ref{eq:Eu-F}).

{\bf Erasure.}
The decoder fails to return the transmitted codeword $\bx=\bx(1)$ if
\begin{equation}
   I(\bx;\by) \le R + \max_{2 \le i \le 2^{NR}} F(I(\bx(i);\by) - R) .
\label{eq:miss-event}
\end{equation}
Denote by $p_\emptyset (p_{\bx})$ the probability of this event.
The event is the disjoint union of events $\calE_1$ and $\calE_2$ below.
The first one is
\begin{equation}
   \calE_1 : \qquad I(\bx;\by) \le R + F(0).
\label{eq:event1}
\end{equation}
Since $F^{-1}$ is increasing, $\calE_1$ is equivalent to $F^{-1}(I(\bx;\by)-R) \le 0$.
The second event is
\begin{eqnarray*}
   \calE_2 : \qquad R + F(0) < I(\bx;\by) & \le &
		R + \max_{2 \le i \le 2^{NR}} F(I(\bx(i);\by) - R) \\
				& = & R + F\left(\max_{2 \le i \le 2^{NR}} I(\bx(i);\by) - R \right)
\end{eqnarray*}
where equality holds because $F$ is nondecreasing. Thus $\calE_2$ is equivalent to
\begin{equation}
   \max_{2 \le i \le 2^{NR}} I(\bx(i);\by) \ge R + F^{-1}(I(\bx;\by) - R) > R .
\label{eq:event2}
\end{equation}

Applying (\ref{eq:I-sp}), we have
\begin{eqnarray}
   Pr[\calE_1] = p_\emptyset^{(1)} (p_{\bx}) & \triangleq & Pr[I(\bx;\by) \le R+F(0)] \nonumber \\
	& \doteq & \exp_2 \left\{ -N E_{sp,N}(R+F(0),p_{\bx},p_{Y|X}) \right\} .
\label{eq:p1}
\end{eqnarray}
Clearly $p_\emptyset (p_{\bx}) \sim p_\emptyset^{(1)} (p_{\bx}) \sim 1$
if $R+F(0) \ge I(p_{\bx},p_{Y|X})$.

Next we have
\begin{eqnarray*}
	Pr[\calE_2] = p_\emptyset^{(2)} (T_{\by|\bx})
    & = & Pr \left[ \max_{2 \le i \le 2^{NR}} I(\bx(i);\by) \ge R + F^{-1}(I(\bx;\by) - R) \right] \\
    & = & Pr \left[ \max_{2 \le i \le 2^{NR}} Z_i \ge F^{-1}(z_1) \right] \\
    & \stackrel{(a)}{\doteq} & 2^{-N F^{-1}(z_1)} \\
	& = & 2^{-N F^{-1}(I(\bx;\by) - R)} 
\end{eqnarray*}
where (a) follows by application of the union bound and (\ref{eq:I-rc}).

Averaging over $\by$, we obtain
\begin{eqnarray}
   p_\emptyset^{(2)} (p_{\bx})
	& \triangleq & \sum_{T_{\by|\bx} \,:\,I(\bx;\by) > R+F(0)}
					Pr[T_{\by|\bx}] \,p_\emptyset^{(2)} (T_{\by|\bx}) \nonumber \\
	& \doteq & \max_{T_{\by|\bx} \,:\,I(\bx;\by) > R+F(0)}
					Pr[T_{\by|\bx}] \,p_\emptyset^{(2)} (T_{\by|\bx}) \nonumber \\
	& \doteq & \max_{p_{\by|\bx} \,:\,I(\bx;\by) > R+F(0)}
				\exp_2 \{ -N[D(p_{\by|\bx} \| p_{Y|X} | p_{\bx}) + F^{-1}(I(\bx;\by) - R)] \} .
\label{eq:p2a}
\end{eqnarray}
For $R+F(0) \le I(p_{\bx},p_{Y|X})$, we have
\[ \max_{p_{\by|\bx} \,:\,I(\bx;\by) \le R+F(0)} \exp_2 \{ -N D(p_{\by|\bx} \| p_{Y|X} | p_{\bx})\}
	\doteq \max_{p_{\by|\bx} \,:\,I(\bx;\by) = R+F(0)} \exp_2 \{ -N D(p_{\by|\bx} \| p_{Y|X} | p_{\bx}) \} ,
\]
hence (\ref{eq:p2a}) may be written more simply as
\begin{eqnarray}
   p_\emptyset^{(2)} (p_{\bx})
	& \doteq & \max_{p_{\by|\bx}} \exp_2 \{ -N[D(p_{\by|\bx} \| p_{Y|X} | p_{\bx}) 
					+ |F^{-1}(I(\bx;\by) - R)|^+] \} 				\nonumber \\
	& = & \exp_2 \{ -N E_{r,|F^{-1}|^+,N}(R,p_{\bx},p_{Y|X}) \} .
\label{eq:p2}
\end{eqnarray}
Since $\calE_1$ and $\calE_2$ are disjoint events, we obtain
\begin{eqnarray*}
   p_\emptyset (p_{\bx}) & = & p_\emptyset^{(1)} (p_{\bx}) + p_\emptyset^{(2)} (p_{\bx}) \\
	& \doteq & \exp_2 \{ -N \min\{ E_{r,|F^{-1}|^+,N}(R,p_{\bx},p_{Y|X}),\,E_{sp,N}(R+F(0),p_{\bx},p_{Y|X})\} \} \\
	& \doteq & \exp_2 \{ -N \min\{ E_{r,|F^{-1}|^+}(R,p_X,p_{Y|X}),\,E_{sp}(R+F(0),p_X,p_{Y|X})\} \}
\end{eqnarray*}
where the last line is due to (\ref{eq:lim-EN}).

The function $F^{-1}(t)$ has a zero-crossing at $t=F(0)$.
Applying (\ref{eq:Esp-ErF-1}), we obtain
\begin{eqnarray*}
   E_{r,|F^{-1}|^+}(R,p_X,p_{Y|X})
	& = & \min_{R'} [E_{sp}(R',p_X,p_{Y|X}) + |F^{-1}(R'-R)|^+] \\
	& \le & E_{sp}(R+F(0),p_X,p_{Y|X}) + 0
\end{eqnarray*}
hence
\[ p_\emptyset (p_{\bx}) \doteq \exp_2 \{ -N E_{r,|F^{-1}|^+}(R,p_X,p_{Y|X}) \} . \]
This proves (\ref{eq:Ee-F}). 
\hfill $\Box$

\Section{Proof of Proposition~\ref{prop:Fopt}}

We first prove (\ref{eq:F-NP-list}).
Recall (\ref{eq:Esp-ErF-1}) and (\ref{eq:FR-property}), restated here for convenience:
\begin{eqnarray*}
   E_{r,F}(R,p_X,\scrW) & = & \min_{R'} [E_{sp}(R',p_X,\scrW) + F(R'-R)] , \\
   E_{sp}(R,p_X,\scrW)  & \equiv &  E_{sp}(R',p_X,\scrW) + F_{R,p_X,\scrW}(R'-R) , \quad \forall R' .
\end{eqnarray*}
Hence
\begin{equation}
  E_{r,F_{R,p_X,\scrW}}(R,p_X,\scrW) = E_{sp}(R,p_X,\scrW) .
\label{eq:ErFR}
\end{equation}

\underline{\em Case~I}: $E_{sp}(R,p_X,\scrW) = \alpha$, i.e., from (\ref{eq:Delta}),
we have $\Delta=0$.
The feasible set ${\calF}^L(R,p_X,\scrW,\alpha)$ defined in (\ref{eq:feasible-F})
takes the form $\{ F~:~ E_{r,F}(R,p_X,\scrW) \ge E_{sp}(R,p_X,\scrW)\}$.
Owing to (\ref{eq:ErFR}) and the monotonicity property of Prop.~\ref{prop:ErF-properties}(ii),
an equivalent representation of ${\calF}^L(R,p_X,\scrW,\alpha)$ is
$\{ F~:~ F \succeq F_{R,p_X,\scrW}\}$.
As indicated below the statement of Prop.~\ref{prop:Fopt},
this implies $F_{R,p_X,\scrW}$ achieves the supremum in (\ref{eq:NP-list}).

\underline{\em Case~II}: $E_{sp}(R,p_X,\scrW) \ne \alpha$.
The derivation parallels that of Case~I. 
An equivalent representation of the constrained set ${\calF}^L(R,p_X,\scrW,\alpha)$
in (\ref{eq:feasible-F}) is $\{ F~:~ F \succeq F^{L*}\}$, where
\begin{eqnarray*}
   F^{L*}(t) & = & F_{R,p_X,\scrW}(t) + \alpha - E_{sp}(R,p_X,\scrW) \\
	& = & F_{R,p_X,\scrW}(t) + \Delta \\
	& = & \alpha - E_{sp}(R+t,p_X,\scrW) . 
\end{eqnarray*}
Hence $F^{L*}$ achieves the maximum in (\ref{eq:NP-list}).

To prove (\ref{eq:F-NP}), we simply observe that if $F^{L*} \preceq F$
for all $F \in {\calF}^L(R,p_X,\scrW,\alpha)$, then $F^* \preceq F$
for all $F \in {\calF}(R,p_X,\scrW,\alpha)$, where $F^*(t) = \max(t,\,F^{L*}(t))$.
\hfill $\Box$

\Section{Proof of Proposition~\ref{prop:Fopt-allR}}

We have
\begin{eqnarray}
   E_\emptyset^L(R,p_X,\scrW,\alpha)
	& \stackrel{(a)}{=} & E_{r,|(F^{L*})^{-1}|^+}(R,p_X,\scrW) \nonumber \\
	& \stackrel{(b)}{=} & \min_{R'}
			[E_{sp}(R',p_X,\scrW) + |(F^{L*})^{-1}|^+ (R'-R)] \nonumber \\
	& \stackrel{(c)}{=} & \min_{R' \ge R+t_{(F^{L*})^{-1}}}
			[E_{sp}(R',p_X,\scrW) + (F^{L*})^{-1}(R'-R)]
\label{eq:ErFinv}
\end{eqnarray}
where equality (a) results from Props.~\ref{prop:EuEe} and \ref{prop:Fopt},
(b) follows from (\ref{eq:Esp-ErF-1}), and (c) from the definition (\ref{eq:tF})
and the fact that the function $E_{sp}(R',p_X,\scrW)$ is nonincreasing in $R'$.
From (\ref{eq:F-NP-list}) and property (P3) in Sec.~\ref{sec:F-MMI}
 we obtain the inverse function
\begin{equation}
   (F^{L*})^{-1}(t) = F_{R,p_X,\scrW}^{-1}(t-\Delta)
\label{eq:F-NP-inv}
\end{equation}
where $F_{R,p_X,\scrW}(t)$ is given in (\ref{eq:FR}). 
Hence $t_{(F^{L*})^{-1}} = \Delta$ and
\begin{equation}
   E_\emptyset^L(R,p_X,\scrW,\alpha) = \min_{R' \ge R+\Delta}
		[\underbrace{E_{sp}(R',p_X,\scrW) + F_{R,p_X,\scrW}^{-1}(R'-R-\Delta)}_{h(R')}] .
\label{eq:ErFinv-caseII}
\end{equation}

By assumption $E_{sp}(R,p_X,\scrW)$ is convex in $R$, and therefore $F_{R,p_X,\scrW}(t)$
is concave. By application of Property (P5) in Sec.~\ref{sec:F-MMI}, the function
$F_{R,p_X,\scrW}^{-1}$ is convex, and thus so is $h(R')$ in (\ref{eq:ErFinv-caseII}).
The derivatives of $F_{R,p_X,\scrW}^{-1}$ and $h$ are respectively given by
\begin{equation}
   (F_{R,p_X,\scrW}^{-1})'(t) = \frac{1}{F_{R,p_X,\scrW}'(t)} = \frac{1}{-E_{sp}'(R+t,p_X,\scrW)}
\label{eq:FR-inv-dev}
\end{equation}
and
\[ h'(R') = E_{sp}'(R',p_X,\scrW) + \frac{1}{-E_{sp}'(R'-\Delta,p_X,\scrW)} . \]

By Prop.~\ref{prop:Fopt} and the definition of $\Delta$ in (\ref{eq:Delta}),
we have $\Ei(R,p_X,\scrW) = \alpha = E_{sp}(R,p_X,\scrW) + \Delta$.
Next we prove the statements (i)---(iv).\\

\noindent
{\bf (i)} $\max(0, R^{conj}(p_X,\scrW) - R) \le \Delta \le I(p_X,\scrW)-R$.

\begin{figure}[hbt]
\begin{center}
\includegraphics[width=13cm]{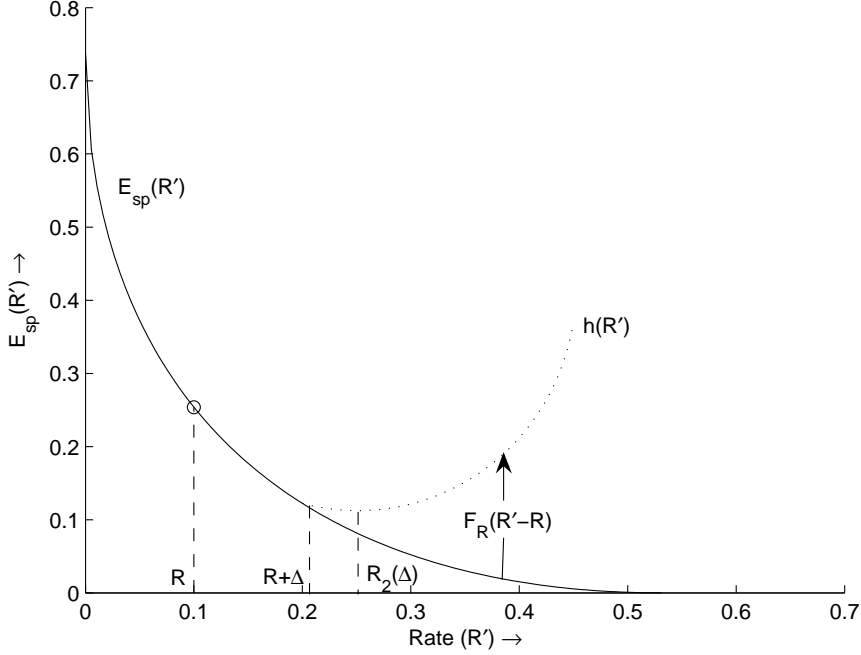}
\end{center}
\caption{Construction and minimization of $h(R')$ for Case (i).}
\label{fig:caseI}
\end{figure}

This case is illustrated in Fig.~\ref{fig:caseI}. We have
\begin{equation}
   R+\Delta \ge \max(R, R^{conj}(p_X,\scrW)) \ge R_{cr}(p_X,\scrW) .
\label{eq:caseI-Rineq}
\end{equation}
Hence
\begin{eqnarray}
   h'(R+\Delta) & = & E_{sp}'(R+\Delta,p_X,\scrW) + \frac{1}{-E_{sp}'(R,p_X,\scrW)} \nonumber \\
	& \ge & \left\{ \begin{array}{ll} E_{sp}'(R,p_X,\scrW) + \frac{1}{-E_{sp}'(R,p_X,\scrW)} \ge 0
					& :~ \mathrm{if~} R \ge R_{cr}(p_X,\scrW) \nonumber \\
			E_{sp}'(R^{conj}(p_X,\scrW),p_X,\scrW) + \frac{1}{-E_{sp}'(R,p_X,\scrW)} = 0
					& :~ \mathrm{if~} R \le R_{cr}(p_X,\scrW)
			\end{array} \right. \nonumber \\
	& \ge & 0 .
\label{eq:h'}
\end{eqnarray}
By convexity of $h(\cdot)$ this implies that $R+\Delta$ minimizes $h(R')$ over $R' \ge R+\Delta$,
and so
\[ E_\emptyset^L(R,p_X,\scrW,\alpha) = h(R+\Delta) = E_{sp}(R+\Delta,p_X,\scrW) . \]
\\

\noindent
{\bf (ii).} 
Due to (\ref{eq:caseI-Rineq}), we have either $R \ge R_{cr}(p_X,\scrW)$ or
$R+\Delta \ge R^{conj}(p_X,\scrW) \ge R$. In both cases,
\[ - E_{sp}'(R,p_X,\scrW) \le \frac{1}{- E_{sp}'(R+\Delta,p_X,\scrW)} . \]
Let $F(t) = \Delta + \lambda |t|^+$
where $\lambda$ is sandwiched by the left and right sides of the above inequality.
We have $t_F=0$ and $F'(t) = \lambda \,\mathds1_{\{t \ge 0\}}$.
The inverse function is $F^{-1}(t) = \frac{1}{\lambda} (t-\Delta)$ for $t \ge \Delta$.
Hence $(F^{-1})'(t) = \frac{1}{\lambda} \,\mathds1_{\{t \ge \Delta\}}$ and $t_{F^{-1}}=\Delta$.
Substituting $F$ and $F^{-1}$ into (\ref{eq:Esp-ErF-1}), we obtain
\begin{eqnarray*}
   E_{r,F}(R,p_X,\scrW) & = & \min_{R' \ge R} [E_{sp}(R',p_X,\scrW) + \Delta + \lambda(R'-R)] , \\
   E_{r,|F^{-1}|^+}(R,p_X,\scrW) & = & \min_{R' \ge R+\Delta} \left[ 
			E_{sp}(R',p_X,\scrW) + \frac{1}{\lambda} (R'-R-\Delta) \right] .
\end{eqnarray*}
Taking derivatives of the bracketed terms with respect to $R'$ and recalling that
\[ \lambda \ge - E_{sp}'(R,p_X,\scrW), \quad \frac{1}{\lambda} \ge - E_{sp}'(R+\Delta,p_X,\scrW) , \]
we observe that these derivatives are nonnegative.
Since $E_{sp}(\cdot,p_X,\scrW)$ is convex, the minima are achieved at $R$ and $R+\Delta$ respectively.

The resulting exponents are $E_{sp}(R,p_X,\scrW)+\Delta$ and $E_{sp}(R+\Delta,p_X,\scrW)$
which coincide with the optimal exponents of (\ref{eq:E-regimeI}). \\

\noindent
{\bf (iii).} $R \le R_{cr}(p_X,\scrW)$ and $0 \le \Delta \le R^{conj}(p_X,\scrW)-R$.

\begin{figure}[hbt]
\begin{center}
\includegraphics[width=13cm]{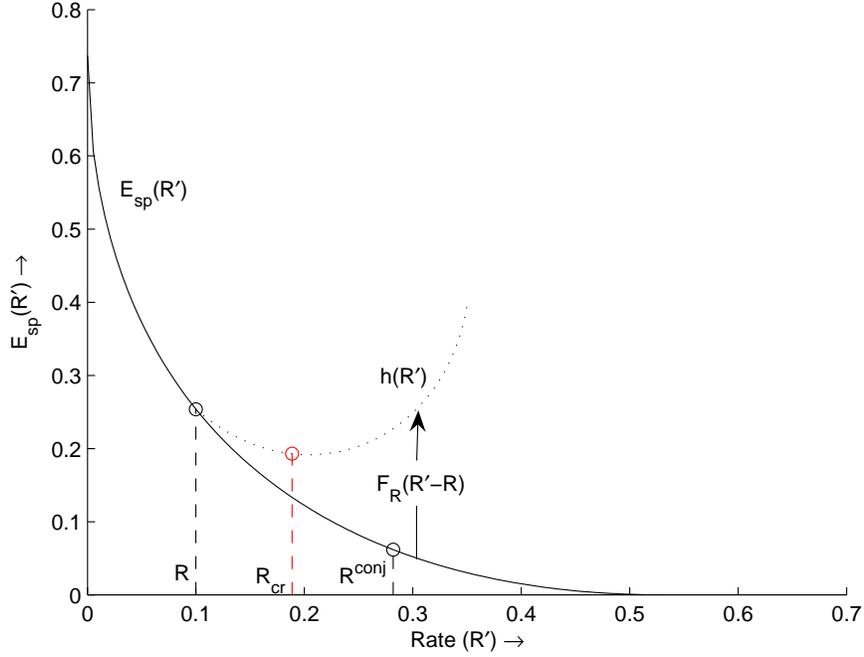}
\end{center}
\caption{Construction and minimization of $h(R')$ for Case (ii), with $\Delta=0$.}
\label{fig:caseII}
\end{figure}

This case is illustrated in Fig.~\ref{fig:caseII} in the case $\Delta=0$.
From (\ref{eq:h'}), we have $h'(R')=0$ if and only if $R'$ and $R'-\Delta$ are conjugate rates.
In this case, using the above assumption on $\Delta$, we have
\begin{equation}
   R \le R'-\Delta = R_1(\Delta) \quad \le R_{cr}(p_X,\scrW) 
		\le \quad R' = R_2(\Delta) \le R^{conj}(p_X,\scrW) .
\label{eq:hmin-caseIIa}
\end{equation}
Hence $R_2(\Delta) = R' \ge R + \Delta$ is feasible for (\ref{eq:ErFinv-caseII})
and minimizes $h(\cdot)$. Substituting $R'$ back into (\ref{eq:ErFinv-caseII}), we obtain
\[ E_\emptyset^L(R,p_X,\scrW,\alpha) = h(R_2(\Delta)) 
	= E_{sp}(R_2(\Delta),p_X,\scrW) + F_{R,p_X,\scrW}^{-1}(R_1(\Delta)-R) \]
which establishes (\ref{eq:E-regimeII}).
	
\noindent
{\bf (iv).} $R \le R_{cr}(p_X,\scrW)$ and $R_\infty(p_X,\scrW) - R \le \Delta \le 0$.

Again we have $h'(R')=0$ if and only if $R'$ and $R'-\Delta$ are conjugate rates.
Then, using the above assumption on $\Delta$, we have
\begin{equation}
   R \le R' = R_1(\Delta) \quad \le R_{cr}(p_X,\scrW) \le \quad R'-\Delta = R_2(\Delta) .
\label{eq:hmin-caseIIb}
\end{equation}
Hence $R_1(\Delta) = R' \ge R + \Delta$ is feasible for (\ref{eq:ErFinv-caseII})
and minimizes $h(\cdot)$. Substituting $R'$ back into (\ref{eq:ErFinv-caseII}), we obtain
\[ E_\emptyset^L(R,p_X,\scrW\alpha) = h(R_1(\Delta)) 
	= E_{sp}(R_1(\Delta),p_X,\scrW) + F_{R,p_X,\scrW}^{-1}(R_2(\Delta)-R) \]
which establishes (\ref{eq:E-regimeIII}).
\hfill $\Box$

\Section{Proof of Lemma~\ref{lem:E'sup}}

First we prove (\ref{eq:E'sp-ineq}). For any $R_0 < R_1$, we have
\begin{eqnarray}
   \int_{R_0}^{R_1} \underline{E}_{sp}'(R,p_X,\scrW) \,dR
   & \stackrel{(a)}{=} & \int_{R_0}^{R_1} \min_{p_{Y|X} \in \scrW} E_{sp}'(R,p_X,p_{Y|X}) \,dR \nonumber \\
   & \le & \min_{p_{Y|X} \in \scrW} \int_{R_0}^{R_1} E_{sp}'(R,p_X,p_{Y|X}) \,dR \nonumber \\
   & = & \min_{p_{Y|X} \in \scrW} [E_{sp}(R_1,p_X,p_{Y|X}) - E_{sp}(R_0,p_X,p_{Y|X})] \nonumber \\
   & \stackrel{(b)}{\le} & E_{sp}(R_1,p_X,p_{Y|X}^*) - E_{sp}(R_0,p_X,p_{Y|X}^*) \nonumber \\
   & = & \min_{p_{Y|X} \in \scrW} E_{sp}(R_1,p_X,p_{Y|X}) - E_{sp}(R_0,p_X,p_{Y|X}^*) \nonumber \\
   & \le & \min_{p_{Y|X} \in \scrW} E_{sp}(R_1,p_X,p_{Y|X})
			- \min_{p_{Y|X} \in \scrW} E_{sp}(R_0,p_X,p_{Y|X}) \nonumber \\
   & = & E_{sp}(R_1,p_X,\scrW) - E_{sp}(R_0,p_X,\scrW) \nonumber \\
   & = & \int_{R_0}^{R_1} E_{sp}'(R_,p_X,\scrW) \,dR
\label{eq:Esp'-ineq}
\end{eqnarray}
where (a) follows from the definition of $\underline{E}_{sp}'$ in (\ref{eq:Esup}),
and we choose $p_{Y|X}^*$ in inequality (b) as the minimizer of $E_{sp}(R_1,p_X,\cdot)$ over $\scrW$.
Since (\ref{eq:Esp'-ineq}) holds for all $R_0 < R_1$, we must have inequality between
the integrands in the left and right sides: $\underline{E}_{sp}'(R,p_X,\scrW) \le E_{sp}'(R_,p_X,\scrW)$.
Moreover, the three inequalities used to derive (\ref{eq:Esp'-ineq}) hold with equality
if the same $p_{Y|X}^{**}$ minimizes $E_{sp}'(R,p_X,\cdot)$ at all rates, and
the same $p_{Y|X}^*$ minimizes $E_{sp}(R,p_X,\cdot)$ at all rates.
We need not (and generally do not) have $p_{Y|X}^{**} = p_{Y|X}^*$. \footnote{
   While $p_{Y|X}^*$ is the noisiest channel in $\scrW$, $p_{Y|X}^{**}$ may be the cleanest channel
   in $\scrW$, as in the BSC example of Sec.~\ref{sec:BSC}.}

Next we prove (\ref{eq:Rconj-ineq}).
By definition of $R^{conj}(p_X, \scrW)$, we have
\begin{eqnarray*}
  \lefteqn{- E_{sp}'(R^{conj}(p_X, \scrW),p_X,\scrW) = \frac{1}{- E_{sp}'(R,p_X,\scrW)} } \\
	& \stackrel{(a)}{\ge} & \frac{1}{- \underline{E}_{sp}'(R,p_X,\scrW)}
	= \min_{p_{Y|X} \in \scrW} \frac{1}{-E_{sp}'(R,p_X,p_{Y|X})}
	= \min_{p_{Y|X} \in \scrW} [ - E_{sp}'(R^{conj}(p_X,p_{Y|X}),p_X,p_{Y|X})] \\
	& \stackrel{(b)}{\ge} & \min_{p_{Y|X} \in \scrW}
		[ - E_{sp}'(\overline{R}^{conj}(p_X, \scrW),p_X,p_{Y|X})] \\ 
	& \stackrel{(c)}{=} & - \underline{E}_{sp}'(\overline{R}^{conj}(p_X, \scrW),p_X,\scrW) \\
	& \stackrel{(d)}{\ge} & - E_{sp}'(\overline{R}^{conj}(p_X, \scrW),p_X,\scrW) .
\end{eqnarray*}
where (a) and (d) are due to (\ref{eq:E'sp-ineq}), (b) to the definition of
$\overline{R}^{conj}(p_X, \scrW)$ in (\ref{eq:Rconj-sup}) and the fact that
$- E_{sp}'(R,p_X,p_{Y|X})$ is a decreasing function of $R$,
and (c) from (\ref{eq:Esup}). Since $- E_{sp}'(R,p_X,\scrW)$ is also a decreasing function
of $R$, we must have $R^{conj}(p_X, \scrW) \le \overline{R}^{conj}(p_X, \scrW)$.
Moreover, the conditions for equality are the same as those for equality in (\ref{eq:E'sp-ineq}).
\hfill $\Box$


\newpage

\begin{thebibliography}{99}

\bibitem{Csiszar81} I. Csisz\'{a}r and J. K\"{o}rner,
	{\em Information Theory: Coding Theory for Discrete Memoryless Systems},
    Academic Press, NY, 1981.

\bibitem{Lapidoth98} A. Lapidoth and P. Narayan, ``Reliable Communication 
	Under Channel Uncertainty,'' {\em IEEE Trans. Information Theory}, 
	Vol.~44, No.~6, pp.~2148---2177, Oct.~1998.

\bibitem{Forney68} G.~D.~Forney,~Jr.,
	``Exponential Error Bounds for Erasure, List, and Decision Feedback Schemes,''
	{\em IEEE Trans. Information Theory}, Vol.~14, No.~2, pp.~206---220,~1968.
	
\bibitem{Telatar94} I.~E.~Telatar and R.~G.~Gallager,
	``New Exponential Upper Bounds to Error and Erasure Probabilities,''
	{\em Proc. ISIT'94}, p.~379, Trondheim, Norway, June 1994.

\bibitem{Merhav07} N.~Merhav and M.~Feder,
	``Minimax Universal Decoding with Erasure Option,''
	{\em IEEE Trans. Information Theory}, Vol.~53, No.~5, pp.~1664---1675, May 2007.
	

\bibitem{Hoeffding65} W.~Hoeffding,
	``Asymptotically Optimal Tests for Multinomial Distributions,''
	{\em Ann. Math. Stat.}, Vol.~36, No.~2, pp.~369---400, 1965.
	
\bibitem{Tusnady77} G. Tusn\'{a}dy, ``On Asymptotically Optimal Tests,''
	{\em Annals of Statistics}, Vol.~5, No.~2, pp.~385---393, 1977.
	
\bibitem{Zeitouni91} O.~Zeitouni and M.~Gutman,
	``On Universal Hypotheses Testing via Large Deviations,''
	{\em IEEE Trans. Information Theory}, Vol.~37, No.~2, pp.~285---290, 1991.

\bibitem{Csiszar07} I. Csisz\'{a}r, {\em personal communication}, Aug.~2007.
	
\bibitem{Feder02} M. Feder and N. Merhav,
	``Universal Composite Hypothesis Testing:
	A Competitive Minimax Approach,''
	{\em IEEE Trans. Information Theory}, 
	Vol.~48, No.~6, pp.~1504---1517, June 2002.

	

\end{thebibliography}
\end{document}